\documentclass[%
  journal=jpcafh
]{achemso}

\setkeys{acs}{
  doi=true
}

\SectionNumbersOn

\usepackage[utf8]{inputenc} 
\usepackage[T1]{fontenc}    
\usepackage{microtype}      
\usepackage{fancyhdr}
\usepackage{graphicx}
\usepackage{physics}
\usepackage{enumitem}
\usepackage{wrapfig}
\usepackage{multirow}
\usepackage{bm}
\usepackage{xcolor}
\usepackage{changepage}   
\definecolor{pigment}{rgb}{0.2, 0.2, 0.6}
\usepackage[%
  colorlinks=true,
  allcolors=pigment
]{hyperref}

\def\kT{k_\mathrm{B}T}                                
\def\d{\mathrm{d}}                                    
\def\e{\mathrm{e}}                                    
\def\dim{\mathrm{dim}}                                
\def\sfP{{\bf P}}                                     
\def\sfQ{{\bf Q}}                                     
\def\sfM{{\bf M}}
\def\bs{{\bf s}}                                      
\def\bx{{\bf x}}                                      
\def\bR{{\bf R}}                                      
\def\bt{{\bm\theta}}                                  
\def\kldiv{D_{\mathrm{KL}}\big(\sfM\|\sfQ\big)}       

\graphicspath{{./}}

\author{Jakub Rydzewski}
\affiliation{Institute of Physics,
Faculty of Physics, Astronomy and Informatics,
Nicolaus Copernicus University,
Grudziadzka 5, 87-100 Torun, Poland}
\email{jr@fizyka.umk.pl}

\author{Omar Valsson}
\affiliation{Max Planck Institute for Polymer Research,
Ackermannweg 10, D-55128 Mainz, Germany}
\email{valsson@mpip-mainz.mpg.de}

\title{Multiscale Reweighted Stochastic Embedding (MRSE): Deep Learning of Collective Variables for Enhanced Sampling}

\begin{document}

\makeatletter
\makeatother

\begin{tocentry}
\centering
\includegraphics[width=\columnwidth]{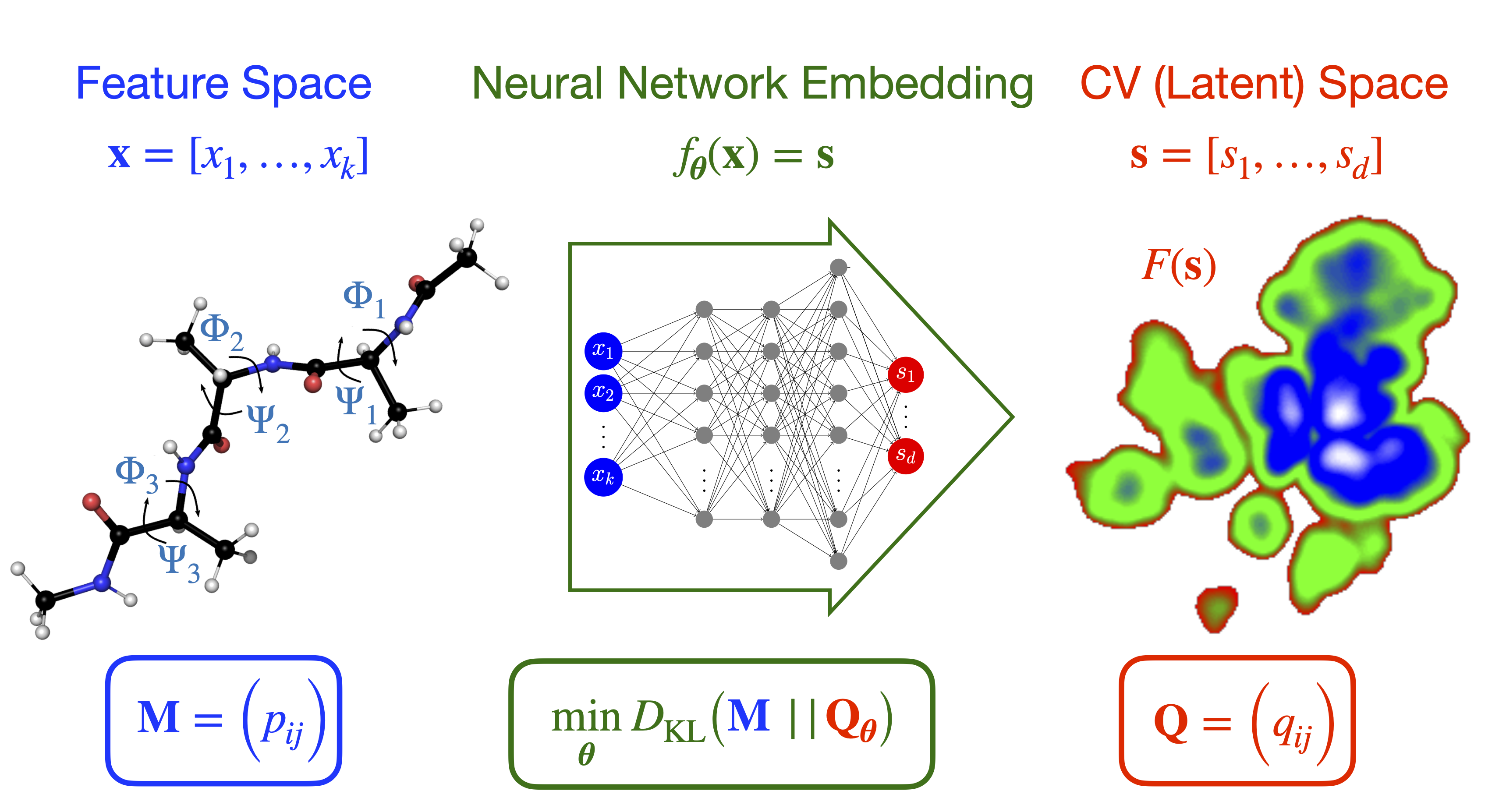}
\end{tocentry}

\newpage

\begin{abstract}
Machine learning methods provide a general framework for automatically finding and representing the essential characteristics of simulation data. This task is particularly crucial in enhanced sampling simulations. There we seek a few generalized degrees of freedom, referred to as collective variables (CVs), to represent and drive the sampling of the free energy landscape. In theory, these CVs should separate different metastable states and correspond to the slow degrees of freedom of the studied physical process. To this aim, we propose a new method that we call multiscale reweighted stochastic embedding (MRSE). Our work builds upon a parametric version of stochastic neighbor embedding. The technique automatically learns CVs that map a high-dimensional feature space to a low-dimensional latent space via a deep neural network. We introduce several new advancements to stochastic neighbor embedding methods that make MRSE especially suitable for enhanced sampling simulations: (1) weight-tempered random sampling as a landmark selection scheme to obtain training data sets that strike a balance between equilibrium representation and capturing important metastable states lying higher in free energy; (2) a multiscale representation of the high-dimensional feature space via a Gaussian mixture probability model; and (3) a reweighting procedure to account for training data from a biased probability distribution. We show that MRSE constructs low-dimensional CVs that can correctly characterize the different metastable states in three model systems: the M\"uller-Brown potential, alanine dipeptide, and alanine tetrapeptide.

\end{abstract}

\newpage

\section{Introduction}
Modeling the long-timescale behavior of complex dynamical systems is a fundamental task in the physical sciences. In principle, molecular dynamics (MD) simulations allow us to probe the spatiotemporal details of molecular processes, but the so-called sampling problem severely limits their usefulness in practice. This sampling problem comes from the fact that a typical free energy landscape consists of many metastable states separated by free energy barriers much higher than the thermal energy $\kT$. Therefore, on the timescale one can simulate, barrier crossings are rare events, and the system remains kinetically trapped in a single metastable state.

One way to alleviate the sampling problem is to employ enhanced sampling methods~\cite{abrams2014enhanced,valsson2016enhancing}. In particular, one class of such methods works by identifying a few critical slow degrees of freedom, commonly referred to as collective variables (CVs), and then enhancing their fluctuations by introducing an external bias potential~\cite{valsson2016enhancing,yang2019enhanced,bussi2020using}. The performance of CV-based enhanced sampling methods depends heavily on the quality of the CVs. Effective CVs should discriminate between the relevant metastable states and include most of the slow degrees of freedom~\cite{noe2017collective}. Typically, the CVs are selected manually by using physical and chemical intuition. Within the enhanced sampling community, numerous generally applicable CVs~\cite{abrams2014enhanced,pietrucci_strategies_2017,rydzewski2017ligand} have been developed and implemented in open-source codes~\cite{Fiorin_2013,tribello2014plumed,Sidky_2018}. However, despite immense progress in devising CVs, it may be far from trivial to find a set of CVs that quantify all the essential characteristics of a molecular system.

Machine learning (ML) techniques, in particular dimensionality reduction or representation learning methods~\cite{murdoch2019definitions,xie2020representation}, provide a possible solution to this problem by automatically finding or constructing the CVs directly from the simulation data~\cite{wang2020machine,noe2020machine,Gkeka2020mlffcgv,sidky2020machine}. Such dimensionality reduction methods typically work in a high-dimensional feature space (e.g., distances, dihedral angles, or more intricate functions~\cite{geiger2013neural,rogal2019neural, musil2021physicsinspired}) instead of directly using the microscopic coordinates, as this is much more efficient. Dimensionality reduction may employ linear or nonlinear transformations, e.g., diffusion map~\cite{coifman2005geometric,coifman2006diffusion,nadler2006diffusion,coifman2008diffusion}, stochastic neighbor embedding (SNE)~\cite{hinton2002stochastic,maaten2008visualizing,maaten2009learning}, sketch-map~\cite{ceriotti2011simplifying,tribello2012using}, and UMAP~\cite{mcinnes2018umap}. In the recent years, there has been a growing interest in performing nonlinear dimensionality reduction with deep neural networks (NNs) to provide parametric embeddings. Inspired by the seminal work of Ma and Dinner~\cite{Ma2005autorc}, several such techniques recently applied to finding CVs include variational autoencoders~\cite{chen2018molecular,hernandez2018variational,ribeiro2018reweighted,chen2018collective}, time-lagged autoencoders~\cite{wehmeyer2018time}, symplectic flows~\cite{li2020neural}, stochastic kinetic embedding~\cite{zhang2018unfolding}, and encoder-map~\cite{Lemke2019EncMap}.

This work proposes a novel technique called multiscale reweighted stochastic embedding (MRSE) that unifies dimensionality reduction via deep NNs and enhanced sampling methods. The method constructs a low-dimensional representation of CVs by learning a parametric embedding from a high-dimensional feature space to a low-dimensional latent space. Our work builds upon various SNE methods~\cite{hinton2002stochastic,maaten2008visualizing,maaten2009learning,van2014accelerating}. We introduce several new aspects to SNE that makes MRSE particularly suitable for enhanced sampling simulations:
\begin{enumerate}
  \item A weight-tempered random sampling as a landmark selection scheme to obtain training data sets that strike a balance between equilibrium representation and capturing important metastable states lying higher in free energy.
  \item Multiscale representation of the high-dimensional feature space via a Gaussian mixture probability model.
  \item Reweighting procedure to account for the sampling of the training data from a biased probability distribution.
\end{enumerate}

We note that the overall objective of our research is to employ MRSE within an enhanced sampling scheme and improve the learned CVs iteratively. However, we focus mainly on the learning procedure for training data from enhanced sampling simulations in this work. Therefore, to eliminate the influence of possible incomplete sampling, we employ idealistic sampling conditions that are generally not achievable in practice~\cite{pant2020statistical}. To gauge the performance of the learning procedure and the quality of the resulting embeddings, we apply MRSE to three model systems (the M\"uller-Brown potential, alanine dipeptide, and alanine tetrapeptide) and provide a thorough analysis of the results.

\section{Methods}
\label{sec:methods}
\subsection{Collective Variable Based Enhanced Sampling}
\label{sec:cv_based_methods}
We start by giving a theoretical background on CV-based enhanced sampling methods. We consider a molecular system, described by microscopic coordinates $\mathbf{R}$ and a potential energy function $U(\mathbf{R})$, which we want to study using MD or Monte Carlo simulations. Without loss of generality, we limit our discussion to the canonical ensemble (NVT). At equilibrium, the microscopic coordinates follow the Boltzmann distribution, $P(\mathbf{R}) = \e^{-\beta U(\mathbf{R})}/\int \d\mathbf{R} \,\e^{-\beta U(\mathbf{R})}$, where $\beta = (k_{\mathrm{B}}T)^{-1}$ is the inverse of the thermal energy.

In CV-based enhanced sampling methods, we identify a small set of coarse-grained order parameters that correspond to the essential slow degrees of freedom, referred to as CVs. The CVs are defined as $\mathbf{s}(\mathbf{R}) = [s_1(\mathbf{R}), s_2(\mathbf{R}), \ldots, s_d(\mathbf{R})]$, where $d$ is the number of CVs (i.e., the dimension of the CV space), and the dependence on $\mathbf{R}$ can be either explicit or implicit. Having defined the CVs, we obtain their equilibrium marginal distribution by integrating out all other degrees of freedom:
\begin{align}
\label{eq:ps}
  P(\mathbf{s}) =
    \int \d\mathbf{R} \, \delta
      \left[
        \mathbf{s} - \mathbf{s}(\mathbf{R})
      \right]
    P(\mathbf{R}),
\end{align}
where $\delta[\cdot]$ is the Dirac delta function. The integral in eq~\ref{eq:ps} is equivalent to $\big< \delta[\mathbf{s}-\mathbf{s(R)}] \big>$, where $\left<\cdot\right>$ denotes an ensemble average. Up to an unimportant constant, the free energy surface (FES) is given by $F(\mathbf{s})= -\beta^{-1} \log P(\mathbf{s})$. In systems plagued by sampling problems, the FES consists of many metastable states separated by free energy barriers much larger than the thermal energy $k_{\mathrm{B}}T$. Therefore, on the timescales we can simulate, the system stays kinetically trapped and is unable to explore the full CV space. In other words, barrier crossings between metastable states are rare events.

CV-based enhanced sampling methods overcome the sampling problem by introducing an external bias potential $V(\mathbf{s}(\mathbf{R}))$ acting in CV space. This leads to sampling according to a biased distribution $P_{V}(\mathbf{R}) = \e^{-\beta \left[U(\mathbf{R})+V(\mathbf{s}(\mathbf{R})) \right]}/\int \d\mathbf{R} \,\e^{-\beta \left[U(\mathbf{R})+V(\mathbf{s}(\mathbf{R})) \right]}$.  We can trace this idea of non-Boltzmann sampling back to the seminal work by Torrie and Valleau published in 1977~\cite{torrie1977nonphysical}. Most CV-based methods adaptively construct the bias potential on-the-fly during the simulation to reduce free energy barriers or even completely flatten them. At convergence, the CVs follow a biased distribution:
\begin{equation}
  P_{V}(\mathbf{s}) =
  \int \d\mathbf{R} \, \delta
    \left[
      \mathbf{s} - \mathbf{s}(\mathbf{R})
    \right]
  P_{V}(\mathbf{R}) =
  \frac{ \e^{ -\beta\left[ F(\mathbf{s}) + V(\mathbf{s}) \right] } }
    {\int\d\mathbf{s} \, \e^{ -\beta\left[ F(\mathbf{s}) + V(\mathbf{s}) \right] } },
\end{equation}
that is easier to sample. CV-based methods differ in how they construct the bias potential and which kind of biased CV sampling they obtain at convergence. A non-exhaustive list of modern CV-based enhanced sampling techniques includes multiple windows umbrella sampling~\cite{Kastner2011umbreallsampling}, adaptive biasing force~\cite{Darve-JCP-2001,Comer2015_TheAdaptiveBiasing,Lesage2016_Smoothed}, Gaussian-mixture umbrella sampling~\cite{Maragakis-JPCB-2009}, metadynamics~\cite{laio2002escaping,barducci2008well,valsson2016enhancing}, variationally enhanced sampling~\cite{valsson2014variational,Valsson2020_VES}, on-the-fly probability-enhanced sampling (OPES)~\cite{Invernizzi2020opus,invernizzi2020unified}, and ATLAS~\cite{gilberti2020atlas}. In the following, we focus on well-tempered metadynamics (WT-MetaD)~\cite{barducci2008well,valsson2016enhancing}. However, we can use MRSE with almost any CV-based enhanced sampling approach.

In WT-MetaD, the time-dependent bias potential is constructed by periodically depositing repulsive Gaussian kernels at the current location in CV space. Based on the previously deposited bias, the Gaussian height is scaled such that it gradually decreases over time~\cite{barducci2008well}. In the long-time limit, the Gaussian height goes to zero. As has been proven~\cite{PhysRevLett.112.240602}, the bias potential at convergence is related to the free energy by:
\begin{equation}
  \label{eq:wt-bias_infty}
  V(\bs,t\to\infty) = -\left( 1-\frac{1}{\gamma} \right) F(\bs),
\end{equation}
and we obtain a so-called well-tempered distribution for the CVs:
\begin{equation}
  \label{eq:wt-pv}
  P_{V}(\mathbf{s}) = \frac{ \left[ P(\mathbf{s}) \right]^{1/\gamma} }
    {\int\d\mathbf{s}\, \left[ P(\mathbf{s}) \right]^{1/\gamma}},
\end{equation}
where $\gamma>1$ is a parameter called bias factor that determines how much we enhance CV fluctuations. The limit $\gamma\to 1$ corresponds to the unbiased ensemble, while the limit $\gamma\to\infty$ corresponds to conventional (non-well-tempered) metadynamics~\cite{laio2002escaping}. If we take the logarithm of both sides of eq~\ref{eq:wt-pv}, we can see that sampling the well-tempered distribution is equivalent to sampling an effective FES, $F_{\gamma}(\bs) = F(\bs)/\gamma$, where the barriers of the original FES are reduced by a factor of $\gamma$. In general, one should select a bias factor $\gamma$ such that effective free energy barriers are on the order of the thermal energy $k_{\mathrm{B}}T$.

Due to the external bias potential, each microscopic configuration $\mathbf{R}$ carries an additional statistical weight $w(\mathbf{R})$ that needs to be taken into account when calculating equilibrium properties. For a static bias potential, the weight is time-independent and given by $w(\mathbf{R})=\e^{\beta V(\mathbf{s}(\mathbf{R}))}$. In WT-MetaD, however, we need to take into account the time-dependence of the bias potential, and thus, the weight is modified in the following way:
\begin{equation}
  \label{eq:weight-wtm}
  w(\mathbf{R},t)=\exp[\beta \tilde{V}( \mathbf{s}(\mathbf{R}), t )],
\end{equation}
where $\tilde{V}(\mathbf{s}(\mathbf{R}),t)=V(\mathbf{s}(\mathbf{R}),t)-c(t)$ is the relative bias potential modified by introducing $c(t)$, a time-dependent constant that can be calculated from the bias potential at time $t$ as~\cite{tiwary_rewt,valsson2016enhancing}:
\begin{equation}
  \label{eq:coft}
  c(t)=\frac{1}{\beta}\log{
  \frac{\int \d\mathbf{s}\,
  \exp\left[
  \frac{\gamma}{\gamma-1} \beta V(\mathbf{s},t)
  \right]}
  {\int \d\mathbf{s}\,
  \exp\left[
  \frac{1}{\gamma-1} \beta V(\mathbf{s},t)
  \right]}}.
\end{equation}
There are also other ways to reweight WT-MetaD simulations~\cite{bonomi_rewt,Branduardi-JCTC-2012,Giberti_2019,Sch_fer_2020}.

In MD simulations, we do not only need to know the values of the CVs but also their derivatives with respect to the microscopic coordinates, $\nabla_{\mathbf{R}} \, \mathbf{s}(\mathbf{R})$. The derivatives are needed to calculate the biasing force $-\nabla_{\mathbf{R}} \, V(\mathbf{s}(\mathbf{R})) = -\partial_{\mathbf{s}} V(\mathbf{s}) \cdot\nabla_{\mathbf{R}} \, \mathbf{s}(\mathbf{R})$. In practice, however, the CVs might not depend directly on $\mathbf{R}$, but rather indirectly through a set of some other input variables (e.g., features). We can even define a CV that is a chain of multiple variables that depend sequentially on each other. In such cases, it is sufficient to know the derivatives of the CVs with respect to the input variables, as we can obtain the total derivatives via the chain rule. In codes implementing CVs and enhanced sampling methods~\cite{Fiorin_2013,tribello2014plumed,Sidky_2018}, like \textsc{plumed}~\cite{tribello2014plumed,plumed-nest}, the handling of the chain rule is done automatically. Thus, when implementing a new CV, we only need to calculate its values and derivatives with respect to the input variables.

Having provided the basics of CV-based enhanced sampling simulations, we now introduce our method for learning CVs.

\subsection{Multiscale Reweighted Stochastic Embedding (MRSE)}
\label{sec:mrse}
The basis of our method is the $t$-distributed variant of stochastic neighbor embedding ($t$-SNE)~\cite{maaten2008visualizing}, a dimensionality reduction algorithm for visualizing high-dimensional data, for instance, generated by unbiased MD simulations~\cite{rydzewski2016machine,zhou2018t,spiwok2020time,fleetwood2021identification}. We introduce here a parametric and multiscale variant of SNE aimed at learning CVs from atomistic simulations. In particular, we focus on using the method within enhanced sampling simulations, where we need to consider biased simulation data. We refer to this method as multiscale reweighted stochastic embedding or MRSE.

We consider a high-dimensional feature space, $\bx=[x_1, \dots, x_k]$, of dimension $k$. The features could be distances, dihedral angles, or some more complex functions~\cite{geiger2013neural,rogal2019neural,musil2021physicsinspired}, which depend on the microscopic coordinates. We introduce a parametric embedding function $f_\bt(\bx)=\bs(\bx)$, that depends on parameters $\bt$, to map from the high-dimensional feature space to the low-dimensional latent space (i.e., the CV space), $\bs=[s_1, \dots, s_d]$, of dimension $d$. From a molecular simulation, we collect $N$ observations (or simply samples) of the features, $[\bx_1, \dots, \bx_N]^T$, that we use as training data. Using these definitions, the problem of finding a low-dimensional set of CVs amounts to using the training data to find an optimal parametrization for the embedding function given a nonlinear ML model. We can then use the embedding as CVs and project any point in feature space to CV space.

In SNE methods, this problem is approached by taking the training data and modeling the pairwise probability distributions for distances in the feature and latent space. To establish the notation, we write the pairwise probability distributions as $\sfM=(p_{ij})$ and $\sfQ=(q_{ij})$, where $1\leq i,j \leq N$, for the feature and the latent space, respectively. For the pairwise probability distribution $\sfM$ ($\sfQ$), the interpretation of a single element $p_{ij}$ ($q_{ij}$) is that higher the value, higher is the probability of picking $\bx_j$ ($\bs_j$) as a neighbor of $\bx_i$ ($\bs_i$). The mapping from the feature space to the latent space is then varied by adjusting the parameters $\bt$ to minimize a loss function that measures the statistical difference between the two pairwise probability distributions. In the following, we explicitly introduce the pairwise probability distributions and the loss function used in MRSE.

\subsubsection{Feature Pairwise Probability Distribution}
\label{sec:feature_distribution}
We model the feature pairwise probability distribution for a pair of samples $\bx_i$ and $\bx_j$ from the training data as a discrete Gaussian mixture. Each term in the mixture is a Gaussian kernel:
\begin{equation}
  K_{\varepsilon_i}(\bx_i,\bx_j)=\exp\left(-\varepsilon_i\|\bx_i-\bx_j\|^2_2\right)
\end{equation}
that is characterized by a scale parameter $\varepsilon_i$ associated to feature sample $\bx_i$. A scale parameter is defined as $\varepsilon_i=1/(2\sigma^2_i)$, where $\sigma_i$ is the standard deviation (i.e., bandwidth) of the Gaussian kernel. Because $\varepsilon_i \neq \varepsilon_j$, the kernels are not symmetric. To measure the distance between data points, we employ the Euclidean distance $\|\cdot\|_2$ as an appropriate metric for representing high-dimensional data on a low-dimensional manifold~\cite{globerson2007euclidean}. Then, a pair $\bx_i$ and $\bx_j$ of points close to each other, as measured by the Euclidean distance, have a high probability of being neighbors.

\begin{figure}[htp]
  \includegraphics[width=0.5\columnwidth]{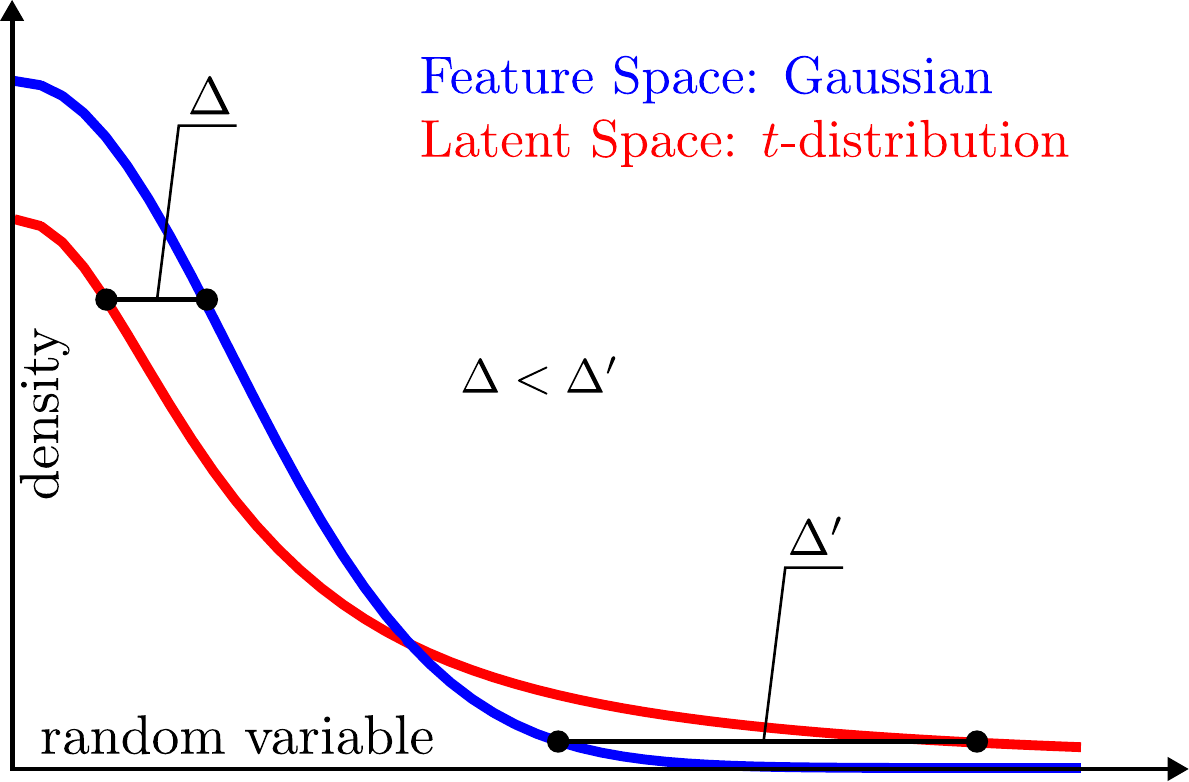}
  \caption{Schematic representation depicting how MRSE (and $t$-SNE) preserves the local structure of high-dimensional data. The pairwise probability distributions are represented by Gaussian kernels in the high-dimensional feature space and by the $t$-distribution kernels in the low-dimensional latent space. The minimization of the Kullback-Leibler divergence between the pairwise probability distributions enforces similar feature samples close to each other and separates dissimilar feature samples in the latent space. As the difference between the distributions fulfills $\Delta'>\Delta$, MRSE is likely to group close-by points into metastable states that are well separated.}
  \label{fig:probabilities}
\end{figure}

For training data obtained from an enhanced sampling simulation, we need to correct the feature pairwise probability distribution because each feature sample $\bx$ has an associated statistical weight $w(\bx)$. To this aim, we introduce a reweighted Gaussian kernel as:
\begin{equation}
  \label{eq:reweighted_kernel}
  \tilde{K}_{\varepsilon_{i}}(\bx_i, \bx_j) =
    r(\bx_i, \bx_j) K_{\varepsilon_{i}}(\bx_i,\bx_j),
\end{equation}
where $r(\bx_i, \bx_j)=\sqrt{w(\bx_i)w(\bx_j)}$ is a pairwise reweighting factor. As noted previously, the exact expression for the weights depends on the enhanced sampling method used. For training data from an unbiased simulation, or if we do not incorporate the weights into the training, all the weights are equal to one and $r(\bx_i, \bx_j) \equiv 1$ for $1 \leq i,j \leq N$.

A reweighted pairwise probability distribution for the feature space is then written as:
\begin{equation}
\label{eq:ppp}
  \sfP=
    \Big(
      p^{\boldsymbol{\varepsilon}}_{ij}
    \Big)_{1\leq i,j \leq N}
    ~~\text{and}~~
  p^{\boldsymbol{\varepsilon}}_{ij}=
    \frac{\tilde{K}_{\varepsilon_{i}}(\bx_i, \bx_j)}
      {\sum_{k} \tilde{K}_{\varepsilon_{i}}(\bx_i, \bx_k)},
\end{equation}
with $p^{\boldsymbol{\varepsilon}}_{ii}=0$. This equation represents the reweighted pairwise probability of features $\bx_i$ and $\bx_j$ for a given set of scale parameters $\boldsymbol{\varepsilon}=[\varepsilon_1,\varepsilon_2, \dots, \varepsilon_N]$, where each scale parameter is assigned to a row of the matrix $\sfP$. The pairwise probabilities $p^{\boldsymbol{\varepsilon}}_{ij}$ are not symmetric due to the different values of the scale parameters ($\varepsilon_i \neq \varepsilon_j$), which is in contrast to $t$-SNE, where the symmetry of the feature pairwise probability distribution is enforced~\cite{maaten2008visualizing}.

As explained in Section~\ref{sec:multiscale_rep} below, the multiscale feature pairwise probability distribution $\sfM$ is written as a mixture of such pairwise probability distributions, each with a different set of scale parameters. In the next section, we describe how to calculate the scale parameters for the probability distribution given by eq~\ref{eq:ppp}.

\subsubsection{Entropy of the Reweighted Feature Probability Distribution}
\label{sec:entropies}
The scale parameters $\boldsymbol{\varepsilon}$ used for the reweighted Gaussian kernels in eq~\ref{eq:ppp} are positive scaling factors that need to be optimized to obtain a proper density estimation of the underlying data. We have that $\varepsilon_i=1/(2\sigma^2_i)$, where $\sigma_i$ is the standard deviation (i.e., bandwidth) of the Gaussian kernel. Therefore, we want a smaller $\sigma_i$ in dense regions and a larger $\sigma_i$ in sparse regions. To achieve this task, we define the Shannon entropy of the $i$th Gaussian probability as:
\begin{equation}
  \label{eq:information_entropy}
  H(\bx_i)=-\sum_j p_{ij}^{\varepsilon_i} \log p_{ij}^{\varepsilon_i},
\end{equation}
where the term $p_{ij}^{\varepsilon_i}$ refers to matrix elements from the $i$th row of $\sfP$ as eq~\ref{eq:information_entropy} is solved for each row independently. We can write $p_{ij}^{\varepsilon_i} = \frac{1}{\bar{p}_i} \tilde{K}_{\varepsilon_{i}}(\bx_i, \bx_j)$ where $\bar{p}_i=\sum_k \tilde{K}_{\varepsilon_{i}}(\bx_i, \bx_k)$ is a row-wise normalization constant.

Inserting $p_{ij}^{\varepsilon_i}$ from eq~\ref{eq:ppp} leads to the following expression:
\begin{align}
  \label{eq:ent}
  H(\bx_i) = \log\bar{p}_i
    &+ \frac{\varepsilon_{i}}{ \bar{p}_i }
       \sum_j \tilde{K}_{\varepsilon_{i}}(\bx_i, \bx_j) \|\bx_i-\bx_j\|_2^2 \nonumber\\
    & \underbrace{ -\frac{1}{ \bar{p}_i }\sum_j \tilde{K}_{\varepsilon_{i}}(\bx_i, \bx_j) \log r(\bx_i,\bx_j) }_{ H_V(\bx_i) },
\end{align}
where $H_V(\bx_i)$ is a correction term due to the reweighting factor $r(\bx_i,\bx_j)$ introduced in eq~\ref{eq:reweighted_kernel}. The reweighting factor is included also in the other two terms through $\tilde{K}_{\varepsilon_{i}}(\bx_i, \bx_j)$. For weights of exponential form, like in WT-MetaD (eq~\ref{eq:weight-wtm}), we have $w(\bx_i)=\e^{\beta V(\bx_i)}$, and the correction term $H_V(\bx_i)$ further reduces to:
\begin{equation}
  \label{eq:entropy_expl}
  H_V(\bx_i)=-\frac{\beta}{2}
  \left(
    \frac{1}{\bar{p}_i}
    \sum_j \tilde{K}_{\varepsilon_{i}}(\bx_i, \bx_j) V(\bx_i) + V(\bx_j)
  \right).
\end{equation}
For the derivation of eq~\ref{eq:ent} and eq~\ref{eq:entropy_expl}, see Section~S1 in the Supporting Information (SI).

For an unbiased simulation, or if we do not incorporate the weights into the training, is $r(\bx_i,\bx_j)\equiv 1$ for $1\leq i,j \leq N$ and the correction term $H_V(\bx_i)$ vanishes. Equation~\ref{eq:ent} then becomes $H(\bx_i) = \log\bar{p}_i + \frac{\varepsilon_{i}}{ \bar{p}_i }\sum_j {K}_{\varepsilon_{i}}(\bx_i, \bx_j) \|\bx_i-\bx_j\|_2^2$.

We use eq~\ref{eq:ent} to define an objective function for an optimization procedure that fits the Gaussian kernel to the data by adjusting the scale parameter so that $H(\bx_i)$ is approximately $\log_2 PP$ (i.e., $\min_{\varepsilon_{i}} \left[H(\bx_i)-\log_2 PP\right]$). Here $PP$ is a model parameter that represents the perplexity of a discrete probability distribution. Perplexity is defined as an exponential of the Shannon entropy, $PP = 2^H$, and measures the quality of predictions for a probability distribution~\cite{cover2006elements}. We can view the perplexity as the effective number of neighbors in a manifold~\cite{maaten2008visualizing,maaten2009learning}. To find the optimal values of the scale parameters, we perform the optimization using a binary search separately for each row of $\sfP$ (eq~\ref{eq:ppp}).

\subsubsection{Multiscale Representation}
\label{sec:multiscale_rep}
As suggested in the work of Hinton and Roweis~\cite{hinton2002stochastic}, the feature probability distribution can be extended to a mixture, as done in refs~\citenum{lee2014multiscale,de2018perplexity,crecchi2020perplexity}. To this aim, for a given value of the perplexity $PP$, we find the optimal set of scale parameters $\boldsymbol{\varepsilon}^{PP}$ using eq~\ref{eq:ent}. We do this for multiple values of the perplexity, $PP_l=2^{L_{PP}-l+1}$, where $l$ goes from 0 to $L_{PP}={\lfloor \log N \rfloor}$-2, and $N$ is the size of the training data set. We then write the probabilities $p_{ij}$ as an average over the different reweighted feature pairwise probability distributions:
\begin{equation}
  \sfM=\Big(p_{ij}\Big)_{1\leq i,j \leq N}
  ~~\text{and}~~
  p_{ij} = \frac{1}{N_{PP}} \sum^{L_{PP}}_{l=0} p^{\boldsymbol{\varepsilon}^{PP_l}}_{ij},
\end{equation}
where $N_{PP}$ is the number of perplexities. Therefore, by taking $p_{ij}$ as a Gaussian mixture over different perplexities, we obtain a multiscale representation of the feature probability distribution $\sfM$, without the need of setting perplexity by the user.

\subsubsection{Latent Pairwise Probability Distribution}
\label{sec:latent_distribution}
A known issue in many dimensionality reduction methods, including SNE, is the so-called ``crowding problem''~\cite{sammon1969nonlinear,hinton2002stochastic}, which is caused partly by the curse of dimensionality~\cite{marimont1979nearest}. In the context of enhanced sampling, the crowding problem would lead to the definition of CVs that inadequately discriminate between metastable states due to highly localized kernel functions in the latent space. As shown in Figure~\ref{fig:probabilities}, if we change from a Gaussian kernel to a more heavy-tailed kernel for the latent space probability distribution, like a $t$-distribution kernel, we enforce that close-by data points are grouped while far-away data points are separated.

Therefore, for the pairwise probability distribution in the latent space, we use a one-dimensional heavy-tailed $t$-distribution, which is the same as in $t$-SNE. We set:
\begin{equation}
\label{eq:Q}
\sfQ=\Big(q_{ij}\Big)_{1\leq i,j \leq N}
~~\text{and}~~
  q_{ij}=\frac{\left(1+\|\bs_i-\bs_j\|^2_2\right)^{-1}}{\sum_{k}%
    \left(1+\|\bs_i-\bs_k\|^2_2\right)^{-1}},
\end{equation}
where $q_{ii}=0$ and the latent variables (i.e., the CVs) are obtained via the embedding function, e.g., $\bs_i = f_\bt(\bx_i)$.

\subsubsection{Minimization of Loss Function}
\label{sec:kl_div}
For the loss function to be minimized during the training procedure, we use the Kullback-Leibler (KL) divergence $\kldiv$ to measure the statistical distance between the pairwise probability distributions $\sfM$ and $\sfQ$~\cite{kullback1951information}. The loss function $L$ for a data batch is defined as:
\begin{align}
\label{eq:kl}
  \kldiv = \frac{1}{N_b}\sum_{i=1}^{N_b}\sum_{\substack{j=1 \\ i \neq j}}^{N_b} p_{ij}\log\left(\frac{p_{ij}}{q_{ij}}\right),
\end{align}
where $\kldiv \geq 0$ with equality only when $\sfM=\sfQ$, and we split the training data into $B$ batches of size $N_b$. We show the derivation of the loss function for the full set of $N$ training data points in Section~S2 in the SI.

\begin{figure}[htp]
  \includegraphics[width=0.4\columnwidth]{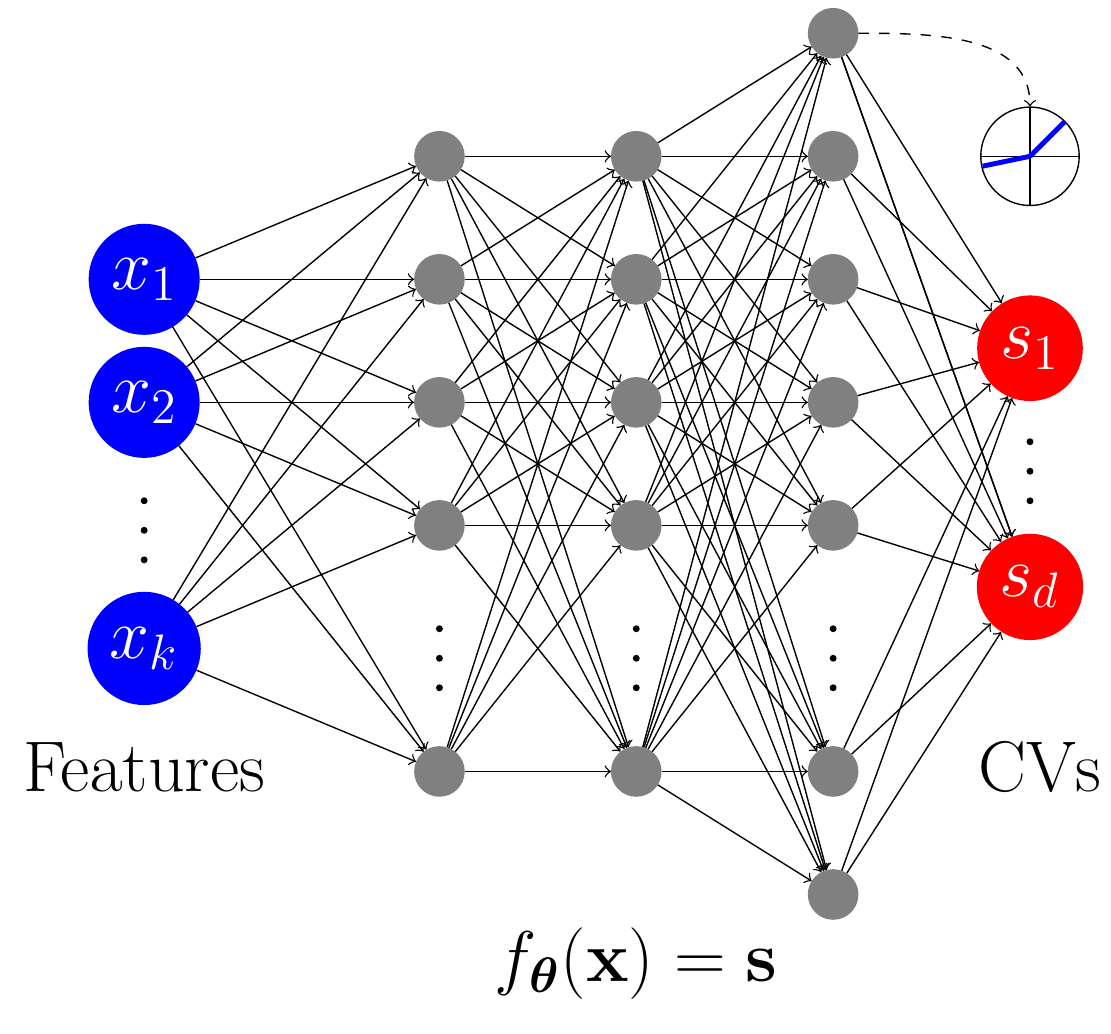}
  \caption{Neural network used to model the parametric embedding function $f_\bt(\bx)$. The input features $\bx$, $\dim(\bx)=k$ are fed into the NN to generate the output CVs $\bs$, $\dim(\bs)=d$. The parameters $\bt$ represent the weights and biases of NN. The input layer is shown in blue, and the output layer is depicted in red. The hidden layers (gray) use dropout and leaky ReLU activations.}
  \label{fig:nn}
\end{figure}

For the parametric embedding function $f_\bt(\bx)$, we employ a deep NN (see Figure~\ref{fig:nn}). After minimizing the loss function, we can use the parametric NN embedding function to project any given point in feature space to the latent space without rerunning the training procedure. Therefore, we can use the embedding as CVs, $\bs(\bx)=f_\bt(\bx)$. The derivatives of $f_\bt(\bx)$ with respect to $\bx$ are obtained using backpropagation. Using the chain rule, we can then calculate the derivatives of $\bs(\bx)$ with respect to the microscopic coordinates $\mathbf{R}$, which is needed to calculate the biasing force in an enhanced sampling simulation.

\subsection{Weight-Tempered Random Sampling of Landmarks}
\label{sec:wtrs}
A common way to reduce the size of a training set is to employ a landmark selection scheme before performing a dimensionality reduction~\cite{ceriotti2013demonstrating,long2019landmark,tribello2019using_a,tribello2019using_b}. The idea is to select a subset of the feature samples (i.e., landmarks) representing the underlying characteristics of the simulation data.

We can achieve this by selecting the landmarks randomly or with some given frequency in an unbiased simulation. If the unbiased simulation has sufficiently sampled phase space or if we use an enhanced sampling method that preserves the equilibrium distribution, like parallel tempering (PT)~\cite{parallel_tempering}, the landmarks represent the equilibrium Boltzmann distribution. However, such a selection of landmarks might give an inadequate representation of transient metastable states lying higher in free energy, as they are rarely observed in unbiased simulations sampling the equilibrium distribution.

For simulation data resulting from an enhanced sampling simulation, we need to account for sampling from a biased distribution when selecting the landmarks. Thus, we take the statistical weights $w(\bR)$ into account within the landmark selection scheme. Ideally, we want the landmarks obtained from the biased simulation to strike a balance between an equilibrium representation and capturing higher-lying metastable states. Inspired by well-tempered farthest-point sampling (WT-FPS)~\cite{ceriotti2013demonstrating} (see Section~S3 in the SI), we achieve this by proposing a simple landmark selection scheme appropriate for enhanced sampling simulations that we call weight-tempered random sampling.

In weight-tempered random sampling, we start by modifying the underlying data density by rescaling the statistical weights of the feature samples as $w(\bR) \rightarrow [w(\bR)]^{1/\alpha}$. Here, $\alpha \geq 1$ is a tempering parameter similar in a spirit to the bias factor $\gamma$ in the well-tempered distribution (eq~\ref{eq:wt-pv}). Next, we randomly sample landmarks according to the rescaled weights. This procedure results in landmarks distributed according to the following probability distribution:
\begin{equation}
  P_\alpha(\bx) =
  \frac
  {\int\d\bR \, \left[w(\bR)\right]^{1/\alpha} \delta\left[\bx-\bx(\bR)\right] P_V(\bR)}
  {\int\d\bR \, \left[w(\bR)\right]^{1/\alpha} P_V(\bR)},
\label{eq:wt-random-sampling}
\end{equation}
which we can rewrite as a biased ensemble average:
\begin{equation}
  P_\alpha(\bx) =
  \frac
  {\Big< [w(\bR)]^{1/\alpha} \delta[\bx - \bx(\bR)] \Big>_V}
  {\Big< [w(\bR)]^{1/\alpha} \Big>_V}.
\label{eq:wt-palpha}
\end{equation}
Similar weight transformations have been used for treating weights degeneracy in importance sampling~\cite{koblents2015population}.

For $\alpha=1$, we recover weighted random sampling~\cite{bortz1975new}, where we sample landmarks according to their unscaled weights $w(\bR)$. As we can see from eq~\ref{eq:wt-random-sampling}, this should, in principle, give an equilibrium representation of landmarks, $P_{\alpha=1}(\bx)=P(\bx)$. By employing $\alpha>1$, we gradually start to ignore the underlying weights when sampling the landmarks and enhance the representation of metastable states lying higher in free energy. In the limit of $\alpha\to\infty$, we ignore the weights (i.e., all are equal to unity) and sample the landmarks randomly so that their distribution should be equal to the biased feature distribution sampled under the influence of the bias potential, $P_{\alpha\to\infty}(\bx)=P_{V}(\bx)$. Therefore, the tempering parameter $\alpha$ allows us to tune the landmark selection between these two limits of equilibrium and biased representation. Using $\alpha>1$ that is not too large, we can obtain a landmark selection that makes a trade-off between an equilibrium representation and capturing higher-lying metastable states.

\begin{figure}[htp]
  \includegraphics[width=0.5\columnwidth]{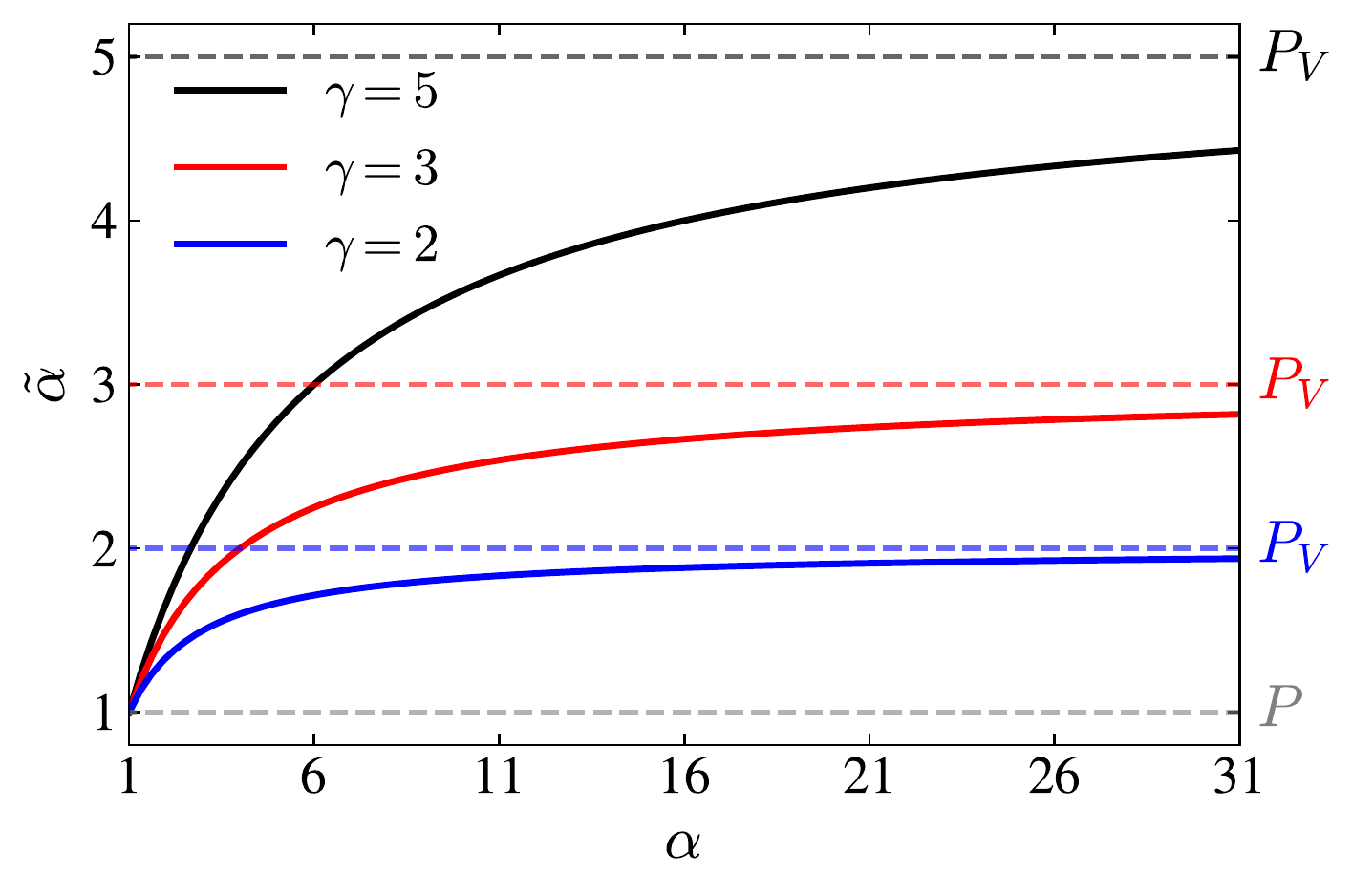}
  \caption{The effective tempering parameter $\tilde{\alpha}$ in the weight-tempered random sampling landmark selection scheme.}
  \label{fig:ala-wt-eff-alpha}
\end{figure}

To understand better the effect of the tempering parameter $\alpha$, we can look at how the landmarks are distributed in the space of the biased CVs for the well-tempered case (eq~\ref{eq:wt-pv}). As shown in Section~S4 in the SI, we obtain:
\begin{equation}
  P_\alpha(\bs) = \frac{ \left[P(\bs)\right]^{1/\tilde{\alpha} }}
    { \int\d\bs\; \left[P(\bs)\right]^{1/\tilde{\alpha} }},
    \label{eq:effective_alpha_pdf}
\end{equation}
where we introduce an effective tempering parameter $\tilde{\alpha}$ as:
\begin{equation}
  \tilde{\alpha}=\left( \frac{1}{\alpha} - \frac{1}{\alpha\gamma} + \frac{1}{\gamma} \right)^{-1} = \frac{\gamma \alpha} {\gamma + \alpha - 1}
  \label{eq:effective_alpha}
\end{equation}
that is unity for $\alpha=1$ and goes to $\gamma$ in the limit $\alpha\to\infty$. Thus, the effect of $\alpha$ is to broaden the CV distribution of the selected landmarks. In Figure~\ref{fig:ala-wt-eff-alpha}, we show how the effective tempering parameter $\tilde{\alpha}$ depends on $\alpha$ for typical bias factor values $\gamma$.

The effect of $\alpha$ on the landmark feature distribution $P_\alpha(\bx)$ is harder to gauge as we cannot write the biased feature distribution $P_V(\bx)$ as a closed-form expression. In particular, for the well-tempered case, $P_V(\bx)$ is not given by $\propto [P(\bx)]^{1/\gamma}$, as the features are generally not fully correlated to the biased CVs~\cite{gil2015enhanced}. The correlation of the features with biased CVs will vary greatly, also within the selected feature set. For example, for features uncorrelated to the biased CVs, the biased distribution is nearly the same as the unbiased distribution. Consequently, the effect of tempering parameter $\alpha$ for a given feature will depend on the correlation with the biased CVs. In Section~\ref{sec:ala1_results}, we will show examples of this issue.

\subsection{Implementation}
\label{sec:implementation}
We implement the MRSE method and the weight-tempered random sampling landmark selection method in an additional module called \texttt{LowLearner} in a development version (2.7.0-dev) of the open-source \textsc{plumed}~\cite{tribello2014plumed,plumed-nest} enhanced sampling plugin. The implementation is available openly at Zenodo~\cite{mrse-dataset} (DOI: \href{https://zenodo.org/record/4756093}{\texttt{10.5281/zenodo.4756093}}) and from the \textsc{plumed} NEST~\cite{plumed-nest} under \texttt{plumID:21.023} at \url{https://www.plumed-nest.org/eggs/21/023/}. We use the LibTorch~\cite{pytorch} library (PyTorch C++ API, git commit \texttt{89d6e88} used to obtain the results in this paper) that allows us to perform immediate execution of dynamic tensor computations with automatic differentiation~\cite{paszke2017automatic}.

\section{Computational Details}
\label{sec:comp_details}
\subsection{Model Systems}
\label{sec:model_systems}
We consider three different model systems to evaluate the performance of the MRSE approach: the M\"uller-Brown Potential, alanine dipeptide, and alanine tetrapeptide. We use WT-MetaD simulations to generate biased simulation data sets used to train the MRSE embeddings for all systems. We also run unbiased simulation data sets for alanine di- and tetrapeptide by performing PT simulations that ensure proper sampling of the equilibrium distribution.

\subsubsection{M\"uller-Brown Potential}
\label{sec:mb_details}
We consider the dynamics of a single particle moving on the two-dimensional M\"uller-Brown potential~\cite{MuellerBrown_1979}, $U(x, y)=\sum_{j} A_{j}\e^{p_{j}(x,y)}$, where $p_{j}(x,y)=a_{j}(x-x_{0,j})^2 + b_{j}(x-x_{0,j})(y-y_{0,j}) + c_{j} (y-y_{0,j})^2$, $x, y$ are the particle coordinates, and $\mathbf{A}, \mathbf{a}, \mathbf{b}, \mathbf{c}, \mathbf{x}_{0}$ and $\mathbf{y}_{0}$ are the parameters of the potential given by $\mathbf{A}=(-40, -20, -34, 3)$, $\mathbf{a}=(-1, -1, 6.5, 0.7)$, $\mathbf{b}=(0,0,11,0.6)$, $\mathbf{c}=(-10,-10,-6.5,-0.7)$, $\mathbf{x}_{0}=(1,0,-0.5,-1$), and $\mathbf{y}_{0}=(0,0.5,1.5,1)$. Note that the $\mathbf{A}$ parameters are not the same as in ref~\citenum{MuellerBrown_1979} as we scale the potential to reduce the height of the barrier by a factor of 5. The FES as a function of the coordinates $x$ and $y$ is given directly by the potential, $F(x,y)=U(x,y)$. We employ rescaled units such that $k_{\mathrm B}=1$. We use the \texttt{pesmd} code from \textsc{plumed}~\cite{tribello2014plumed,plumed-nest} to simulate the system at a temperature of $T=1$ using a Langevin thermostat~\cite{bussi_accurate_2007} with a friction coefficient of 10 and employ a time step of 0.005. At this temperature, the potential has a barrier of around 20 $k_{\mathrm B}T$ between its two states and thus is a rare event system.

For the WT-MetaD simulations, we take $x$ and $y$ as CVs. We use different bias factors values (3, 4, 5, and 7), an initial Gaussian height of 1.2, a Gaussian width of 0.1 for both CVs, and deposit Gaussians every 200 steps. We calculate $c(t)$ (eq~\ref{eq:coft}), needed for the weights, every time a Gaussian is added using a grid of $500^2$ over the domain $[-5,5]^2$. We run the WT-MetaD simulations for a total time of $2 \times 10^7$ steps. We skip the first 20\% of the runs (up to step $4 \times 10^6$) to ensure that we avoid the period at the beginning of the simulations where the weights might be unreliable due to rapid changes of the bias potential. For the remaining part, we normalize the weights such that they lie in the range 0 to 1 to avoid numerical issues.

We employ features saved every 1600 steps for the landmark selection data sets, yielding a total of $10^4$ samples. From these data sets, we then use weight-tempered random sampling with $\alpha=2$ to select 2000 landmarks that we use as training data to generate the MRSE embeddings.

For the embeddings, we use the coordinates $x$ and $y$ as input features ($k=2$), while the number of output CVs is also 2 ($d=2$). We do not standardize or preprocess the input features.

\subsubsection{Alanine Dipeptide}
\label{sec:ala1_details}
We perform the alanine dipeptide (Ace-Ala-Nme) simulations using the \textsc{gromacs} 2019.2 code~\cite{gromacs} patched with a development version of the \textsc{plumed} plugin~\cite{tribello2014plumed,plumed-nest}. We use the Amber99-SB force field~\cite{Hornak2006a}, and a time step of 2 fs. We perform the simulations in the canonical ensemble using the stochastic velocity rescaling thermostat~\cite{bussi2007canonical} with a relaxation time of 0.1 fs. We constrain hydrogen bonds using LINCS~\cite{hess2008p}. The simulations are performed in vacuum without periodic boundary conditions. We employ no cut-offs for electrostatic and non-bonded van der Waals interactions.

We employ 4 replicas with temperatures distributed geometrically in the range 300 K to 800 K (300.0 K, 416.0 K, 576.9 K, 800.0 K) for the PT simulation. We attempt exchanges between neighboring replicas every 10 ps. We run the PT simulation for 100 ns per replica. We only use the 300 K replica for analysis.

We perform the WT-MetaD simulations at 300 K using the backbone dihedral angles $\Phi$ and $\Psi$ as CVs and employ different values for the bias factor (2, 3, 5, and 10). We use an initial Gaussian height of 1.2 kJ/mol, a Gaussian width of 0.2 rad for both CVs, and deposit Gaussians every 1 ps.  We calculate $c(t)$ (eq~\ref{eq:coft}) every time a Gaussian is added (i.e., every 1 ps) employing a grid of $500^2$ over the domain $[-\pi,\pi]^2$. We run the WT-MetaD simulations for 100 ns. We skip the first 20 ns of the runs (i.e., first 20\%) to ensure that we avoid the period at the beginning of the simulations where the weights might be unreliable due to rapid changes in the bias potential. For the remaining part, we normalize the weights such that they lie in the range 0 to 1 to avoid numerical issues.

For the landmark selection data sets, we employ features saved every 1 ps, which results in data sets of $8 \times 10^4$ and $1 \times 10^5$ samples for the WT-MetaD and PT simulations, respectively. We select 4000 landmarks for the training from these data sets, using weighted random sampling for the PT simulation and weight-tempered random sampling for the WT-MetaD simulations ($\alpha=2$ unless otherwise specified).

For the embeddings, we use 21 heavy atoms pairwise distances as input features ($k=21$) and the number of output CVs as 2 ($d=2$). To obtain an impartial selection of features, we start with all 45 heavy atoms pairwise distances. Then, to avoid unimportant features, we automatically check for low variance features and remove all distances with a variance below $2 \times 10^{-4}$ nm$^{2}$ from the training set (see Section~S9 in the SI). This procedure removes 24 distances and leaves 21 distances for the embeddings (both training and projections). We standardize remaining distances individually such that their mean is zero and their standard deviation is one.

\subsubsection{Alanine Tetrapeptide}
\label{sec:ala3_details}
We perform simulations of alanine tetrapeptide (Ace-Ala$_3$-Nme) in vacuum using the \textsc{gromacs} 2019.2 code~\cite{gromacs} and a development version of the \textsc{plumed} plugin~\cite{tribello2014plumed,plumed-nest}. We use the same MD setup and parameters as for the alanine dipeptide system, e.g., the Amber99-SB force field~\cite{Hornak2006a}, see Section~\ref{sec:ala1_details} for further details.

For the PT simulation, we employ 8 replicas with temperatures ranging from 300 K to 1000 K according to a geometric distribution (300.0 K, 356.4 K, 424.3 K, 502.6 K, 596.9 K, 708.9 K, 842.0 K, 1000.0 K). We attempt exchanges between neighboring replicas every 10 ps. We simulate each replica for 100 ns. We only use the 300 K replica for analysis.

We perform the WT-MetaD simulation at 300 K using the backbone dihedral angles $\Phi_1$, $\Phi_2$, and $\Phi_3$ as CVs and a bias factor of 5. We use an initial Gaussian height of 1.2 kJ/mol, a Gaussian width of 0.2 rad, and deposit Gaussians every 1 ps. We run the WT-MetaD simulation for 200 ns. We calculate $c(t)$ every 50 ps using a grid of $200^3$ over the domain $[-\pi,\pi]^3$. We skip the first 40 ns of the run (i.e., first 20\%) to ensure that we avoid the period at the beginning of the simulation where the weights are not equilibrated. We normalize the weights such that they lie in the range 0 to 1.

For the landmark selection data sets, we employ features saved every 2 ps for the WT-MetaD simulation and every 1 ps for the PT simulation. This results in data sets of $8 \times 10^4$ and $1 \times 10^5$ samples for the WT-MetaD and PT simulations, respectively. We select 4000 landmarks for the training from these data sets, using weighted random sampling for the PT simulation and weight-tempered random sampling with $\alpha=2$ for the WT-MetaD simulations.

For the embeddings, we use sines and cosines of the dihedral angles $(\Phi_1,\Psi_1,\Phi_2,\Psi_2,\Phi_3,\Psi_3)$ as input features ($k=12$), and the number of output CVs is 2 ($d=2$). We do not standardize or preprocess the input features further.

\subsection{Neural Network Architecture}
\label{sec:nn_architecture}
For the NN, we use the same size and number of layers as in the work of van der Maaten and Hinton~\cite{maaten2009learning,hinton2006reducing}. The NN consists of an input layer with a size equal to the dimension of the feature space $k$, followed by three hidden layers of sizes $h_1=500$, $h_2=500$, and $h_3=2000$, and an output layer with a size equal to the dimension of the latent space $d$.

To allow for any output value, we do not wrap the output layer within an activation function. Moreover, for all hidden layers, we employ leaky rectified linear units (leaky ReLU)~\cite{maas2013rectifier} with a leaky parameter set to $0.2$. Each hidden layer is followed by a dropout layer~\cite{dropout} (dropout probability $p=0.1$). For the details regarding the architecture of NNs, see Table~\ref{tab:hyperparams}.

\subsection{Training Procedure}
\label{sec:training_details}
We shuffle the training data sets and divide them into batches of size 500. We initialize all trainable weights of the NNs with the Glorot normal scheme~\cite{glorot2010understanding} using the gain value calculated for leaky ReLU. The bias parameters of the NNs are initialized with 0.005.

We minimize the loss function given by eq~\ref{eq:kl} using the Adam optimizer~\cite{kingma2014adam} with AMSGrad~\cite{Reddi2019}, where we use learning rate $\eta=10^{-3}$, and momenta $\beta_1=0.9$ and $\beta_2=0.999$. We also employ a standard L2 regularization term on the trainable network parameters in the form of weight decay set to $10^{-4}$. We perform the training for 100 epochs in all cases. The loss function learning curves for the systems considered here are shown in Section~S7 in the SI.

We report all hyperparameters used to obtain the results in this work in Table~\ref{tab:hyperparams}. For reproducibility purposes, we also list the random seeds used while launching the training runs (the seed affects both the landmark selection and the shuffling of the landmarks during the training).

\begin{table*}[htp!]
  \centering\footnotesize
  \caption{Hyperparameters used to obtain the results reported in this paper.}
  \begin{tabular}{|l|l|l|l|}
  \hline
  Hyperparameter      & M\"uller-Brown         & Alanine dipeptide       & Alanine tetrapeptide      \\
  \hline
  Features            & $x$ and $y$            & Heavy atom distances    & Dihedral angles (cos/sin) \\
  NN architecture     & [2, 500, 500, 2000, 2] & [21, 500, 500, 2000, 2] & [12, 500, 500, 2000, 2]   \\
  Optimizer           & Adam (AMSGrad)         & Adam (AMSGrad)          & Adam (AMSGrad)            \\
  Number of landmarks & $N=2000$               & $N=4000$                & $N=4000$                  \\
  Batch size          & $N_b=500$              & $N_b=500$               & $N_b=500$                 \\
  Training iterations & 100                    & 100                     & 100                       \\
  Learning rate       & $\eta=10^{-3}$         & $\eta=10^{-3}$          & $\eta=10^{-3}$            \\
  Seed                & 111                    & 111 (SI: 222, 333)      & 111                       \\
  Leaky parameter     & 0.2                    & 0.2                     & 0.2                       \\
  Dropout             & $p=0.1$                & $p=0.1$                 & $p=0.1$                   \\
  Weight decay        & $10^{-4}$              & $10^{-4}$               & $10^{-4}$                 \\
  $\beta_1,\beta_2$   & 0.9 and 0.999          & 0.9 and 0.999           & 0.9 and 0.999             \\
  \hline
  \end{tabular}
  \label{tab:hyperparams}
\end{table*}

\subsection{Kernel Density Estimation}
\label{sec:kde}
We calculate FESs for the trained MRSE embeddings using kernel density estimation (KDE) with Gaussian kernels. We employ a grid of $200^2$ for the FES figures. We choose the bandwidths for each simulation data set by first estimating them using Silverman's rule and then adjusting the bandwidths by comparing the KDE FES to an FES obtained with a discrete histogram. We show a representative comparison between KDE and discrete FESs in Section~S6 in the SI. We employ reweighting for FESs from WT-MetaD simulation data where we weigh each Gaussian KDE kernel by the statistical weight $w(\mathbf{R})$ of the given data point.

\subsection{Data Availability}
\label{sec:data}
The data supporting the results of this study are openly available at Zenodo~\cite{mrse-dataset} (DOI: \href{https://zenodo.org/record/4756093}{\texttt{10.5281/zenodo.4756093}}). \textsc{plumed} input files and scripts required to replicate the results presented in the main text are available from the \textsc{plumed} NEST~\cite{plumed-nest} under \texttt{plumID:21.023} at \url{https://www.plumed-nest.org/eggs/21/023/}.

\section{Results}
\label{sec:results}
\subsection{M\"uller-Brown Potential}
\label{sec:mb_results}
\begin{figure}[htp]
  \includegraphics[width=0.8\columnwidth]{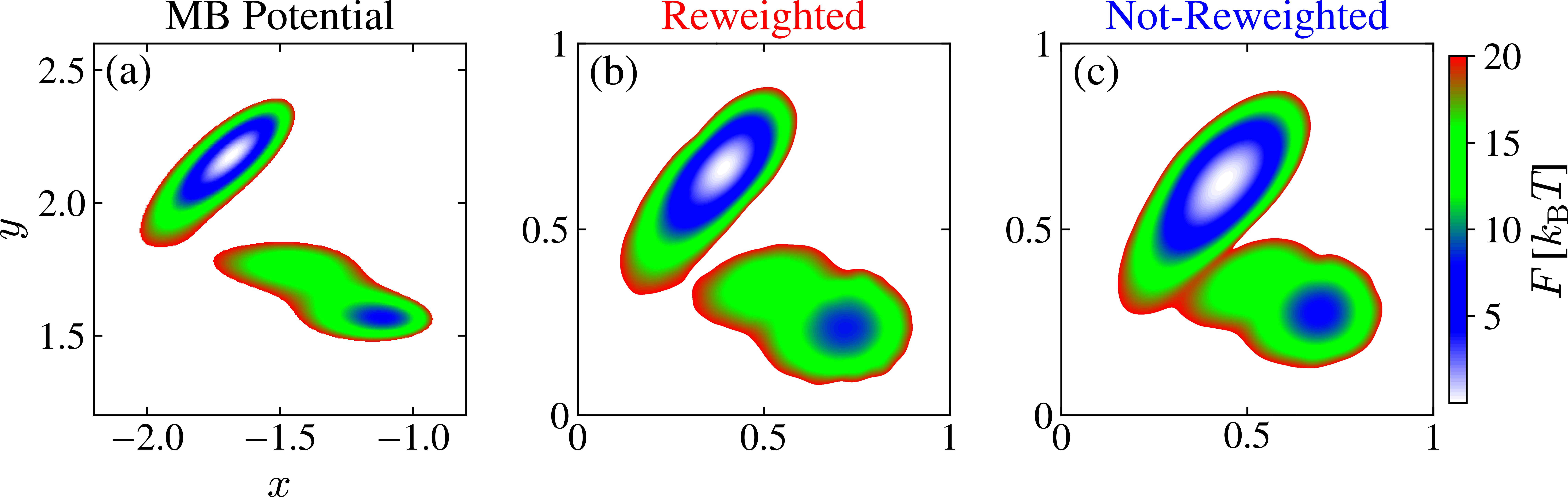}
  \caption{Results for the M\"uller-Brown potential. FESs for MRSE embeddings obtained from the WT-MetaD simulation ($\gamma=5$). We show MRSE embeddings obtained with (b) and without (c) incorporating weights into the training via a reweighted feature pairwise probability distribution (see eq~\ref{eq:reweighted_kernel}). The units for the MRSE embeddings are arbitrary and only shown as a visual guide. To facilitate comparison, we post-process the MRSE embeddings using the Procrustes algorithm to find an optimal rotation that best aligns with the original coordinates $x$ and $y$, see text.}
  \label{fig:mb-emb}
\end{figure}

We start by considering a single particle moving on the two-dimensional M\"uller-Brown potential shown in Figure~\ref{fig:mb-emb}(a). We use this system as a simple test to check if the MRSE method can preserve the topography of the FES in the absence of any dimensionality reduction when performing a mapping with a relatively large NN.

We train the MRSE embeddings on simulation data sets obtained from WT-MetaD simulations using the coordinates $x$ and $y$ as CVs. Here, we show only the results obtained with bias factor $\gamma=5$, while the results for other values are shown in Section~S8 in the SI. The MRSE embeddings can be freely rotated and overall rotation is largely determined by the random seed used to generate the embeddings. Therefore, to facilitate comparison, we show here results obtained using the Procrustes algorithm to find an optimal rotation of the MRSE embeddings that best aligns with the original coordinates $x$ and $y$. The original non-rotated embeddings are shown in Section~S8 in the SI. We present the FESs obtained with the MRSE embeddings in Figure~\ref{fig:mb-emb}(b-c). We can see that the embeddings preserve the topography of the FESs very well and demonstrate a fine separation of metastable states, both when we incorporate the weights into the training through eq~\ref{eq:reweighted_kernel} (panel b), and when we do not (panel c).
\begin{figure}[htp]
  \includegraphics[width=0.4\columnwidth]{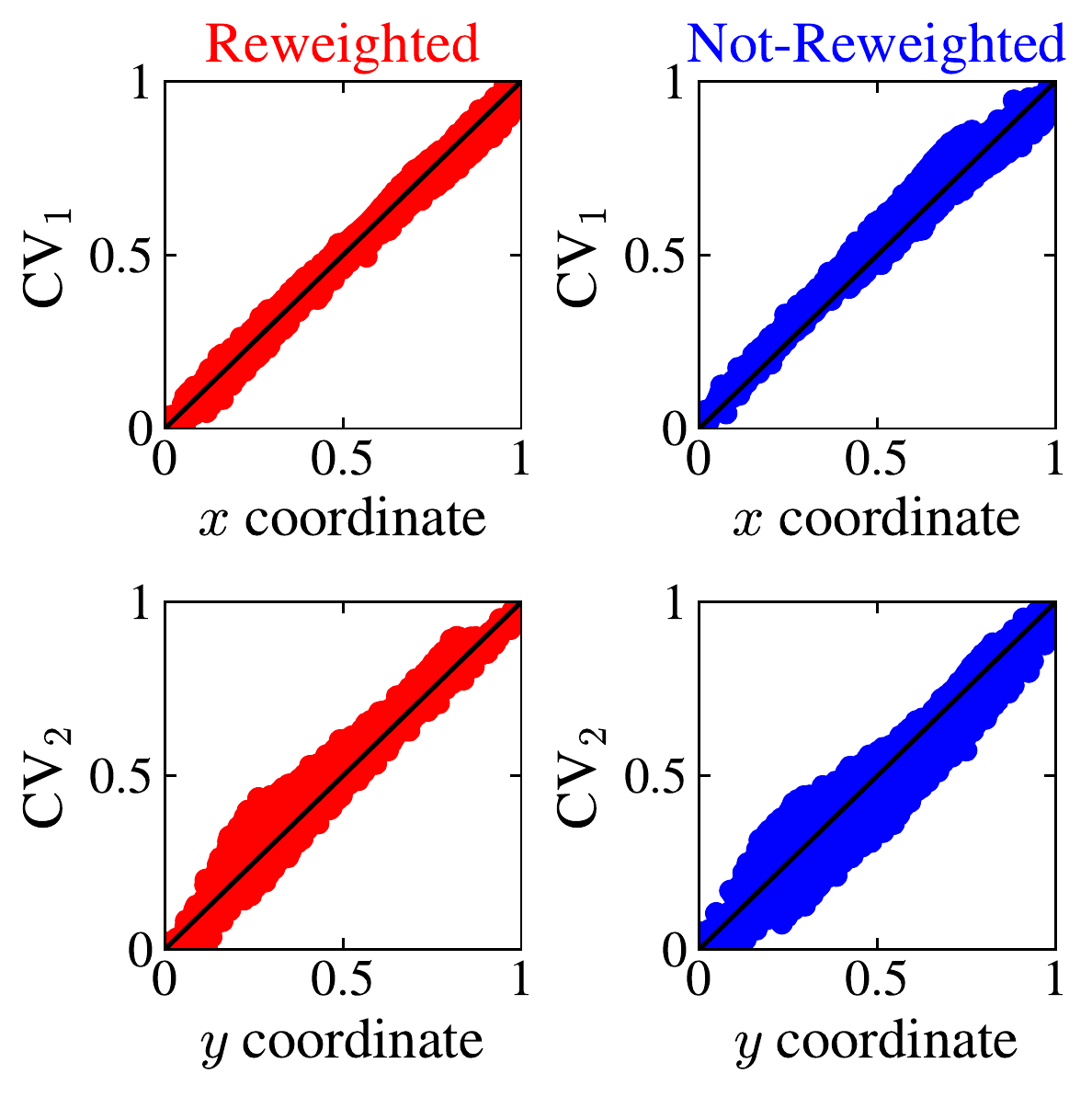}
  \caption{Results for the M\"uller-Brown potential. We show how the MRSE embeddings map the coordinates $x$ and $y$ by plotting the normalized coordinates $x$ and $y$ versus the normalized MRSE CVs. The MRSE embeddings are trained using data from a WT-MetaD simulation with $\gamma=5$, and obtained with (red) and without (blue) incorporating weights into the training via a reweighted feature pairwise probability distribution (see eq~\ref{eq:reweighted_kernel}). To facilitate comparison, we post-process the MRSE embeddings using the Procrustes algorithm to find an optimal rotation that best aligns with the original coordinates $x$ and $y$, see text.}
  \label{fig:mb-vs}
\end{figure}

To quantify the difference between the $x$ and $y$ coordinates and the CVs found by MRSE, we normalize all coordinates and plot CV$_1$ as a function of $x$ and CV$_2$ as a function of $y$. In Figure~\ref{fig:mb-vs}, we can see that the points lie along the identity line, which shows that both MRSE embeddings preserve well the original coordinates of the MB system. In other words, the embeddings maintain the normalized distances between points. We analyze this aspect in a detailed manner for a high-dimensional set of features in Section~\ref{sec:ala1_results}.

\subsection{Alanine Dipeptide}
\label{sec:ala1_results}
\begin{figure}[htp]
  \includegraphics[width=0.5\linewidth]{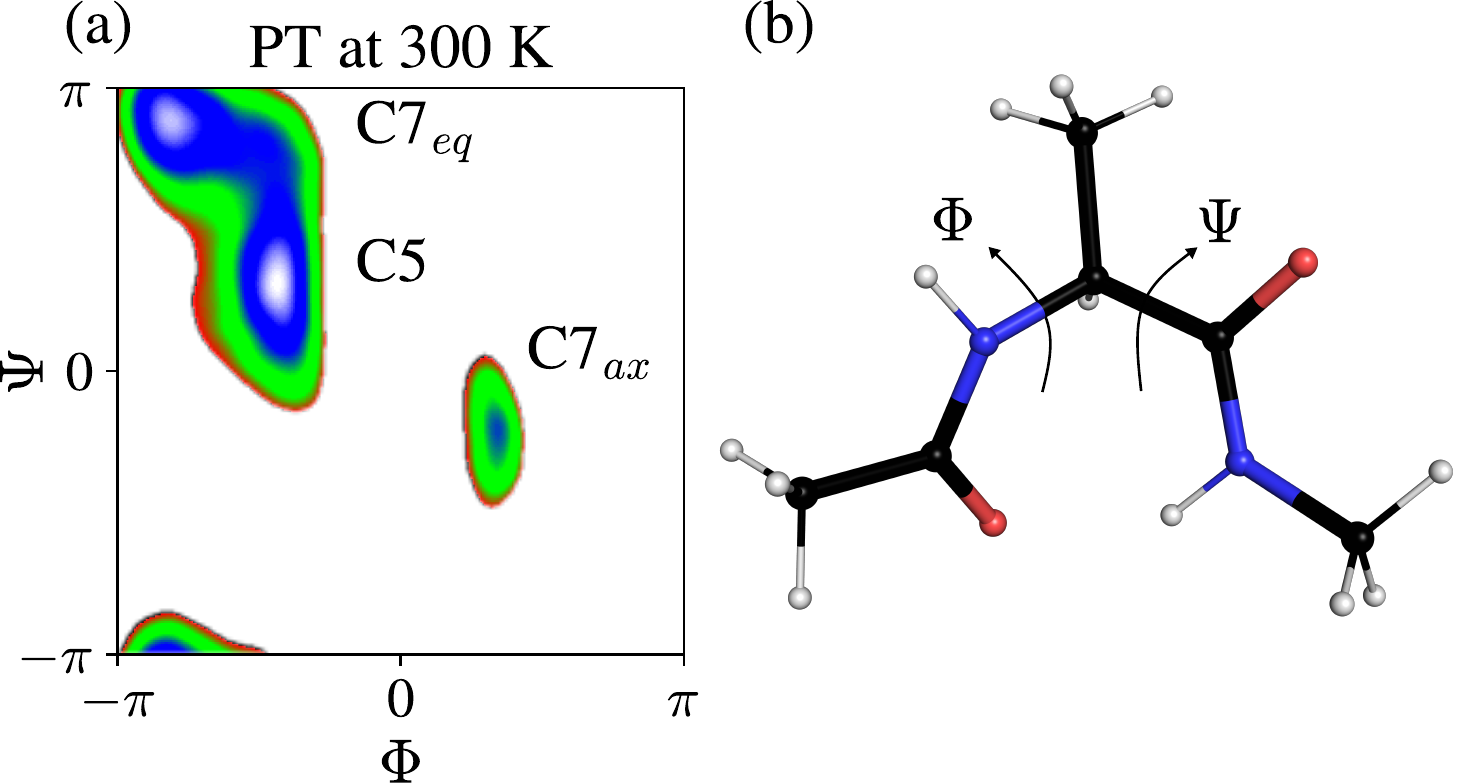}
  \caption{Results for alanine dipeptide in vacuum at 300 K. (a) The free energy landscape $F(\Phi,\Psi)$ from the PT simulation. The metastable states C$7_{\mathrm{eq}}$, C5, and C$7_{\mathrm{ax}}$ are shown. (b) The molecular structure of alanine dipeptide with the dihedral angles $\Phi$ and $\Psi$ indicated.}
  \label{fig:ala-mol}
\end{figure}

Next, we consider alanine dipeptide in vacuum, a small system often used to benchmark free energy and enhanced sampling methods. The free energy landscape of the system is described by the backbone $(\Phi,\Psi)$ dihedral angles. Generally, the $(\Phi,\Psi)$ angles are taken as CVs for biasing, as we do here to generate the training data set. However, for this particular setup in vacuum, it is sufficient to bias $\Phi$ to drive the sampling between states as $\Psi$ is a fast CV compared to $\Phi$. We can see in Figure~\ref{fig:ala-mol} that three metastable states characterize the FES. The C$7_{\mathrm{eq}}$ and C$5$ states are separated only by a small barrier of around 1--2 $k_{\mathrm{B}}T$, so transitions between these two states are frequent. The C$7_{\mathrm{ax}}$ state lies higher in free energy (i.e., is less probable to sample), and is separated by a high barrier of around 14 $k_{\mathrm{B}}T$ from the other two states, so transitions from C$7_{\mathrm{eq}}$/C$5$ to C$7_{\mathrm{ax}}$ are rare.

For the MRSE embeddings, we do not use the $(\Phi,\Psi)$ angles as input features, but rather a set of 21 heavy atom pairwise distances that we impartially select as described in Section~\ref{sec:ala1_details}. Using only the pairwise distances as input features makes the exercise of learning CVs more challenging as the $\Phi$ and $\Psi$ angles cannot be represented as linear combinations of the interatomic distances. We can assess the quality of our results by examining how well the MRSE embeddings preserve the topography of the FES on local and global scales. However, before presenting the MRSE embeddings, let us consider the landmark selection, which we find crucial to our protocol to construct embeddings accurately.

As discussed in Section~\ref{sec:wtrs}, we need to have a landmark selection scheme that takes into account the weights of the configurations and gives a balanced selection that ideally is close to the equilibrium distribution but represents all metastable states of the system, also the higher-lying ones. We devise for this task a method called weight-tempered random sampling. This method has a tempering parameter $\alpha$ that allows us to interpolate between an equilibrium and a biased representation of landmarks (see eq~\ref{eq:wt-random-sampling}).

\begin{figure}[htp]
  \includegraphics[width=0.4\columnwidth]{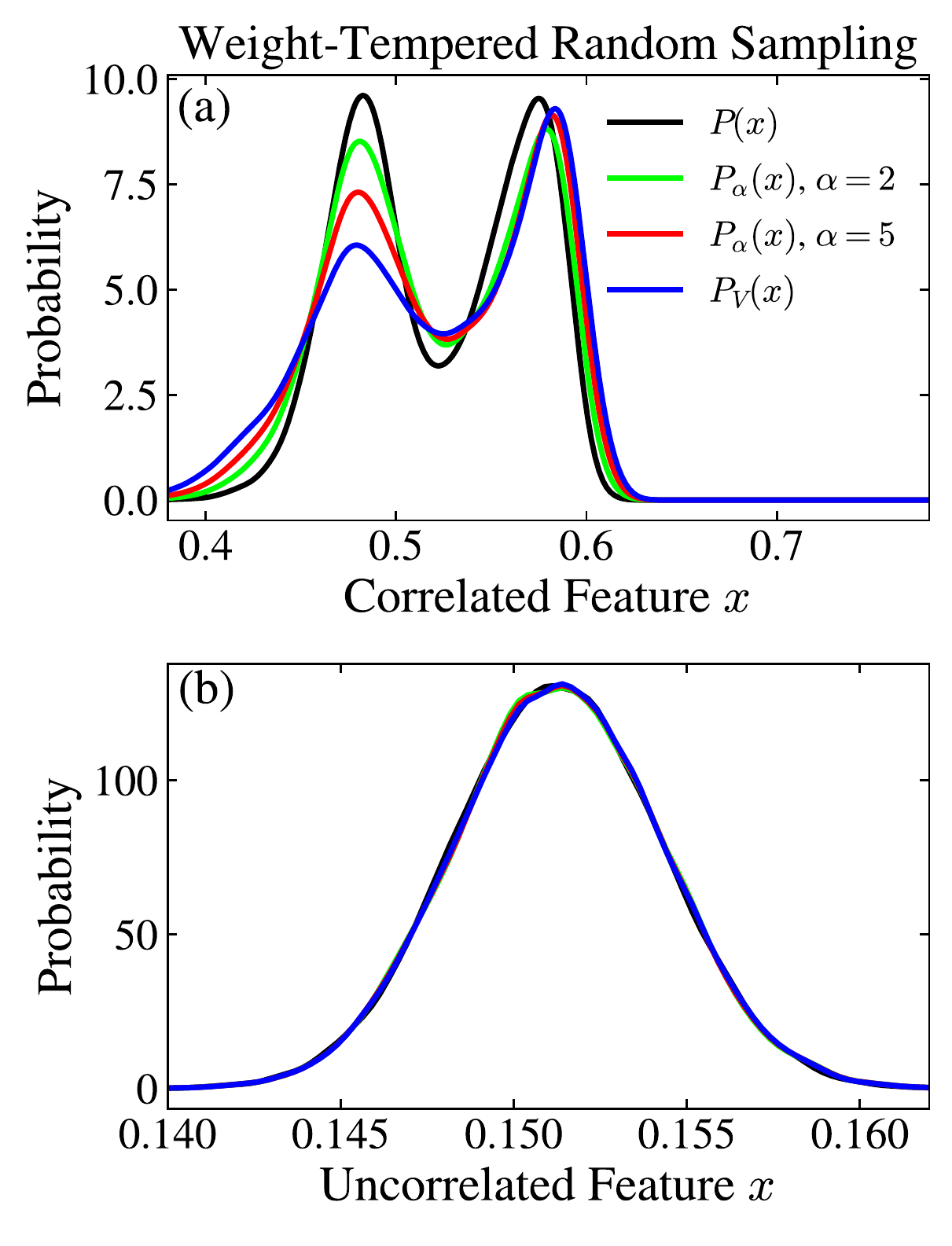}
  \caption{Results for alanine dipeptide in vacuum at 300 K. The effect of the tempering parameter $\alpha$ in the weight-tempered random sampling landmark selection scheme for a WT-MetaD simulation ($\gamma=5$) biasing $(\Phi, \Psi)$. Marginal landmark distributions for two examples of features (i.e., heavy atom distances)) from the feature set that are (a) correlated and (b) uncorrelated with the biased CVs. The units are nm.}
  \label{fig:ala-wt-sampling}
\end{figure}

The effect of the tempering parameter $\alpha$ on the landmark feature distribution $P_\alpha(\bx)$ will depend on the correlation of the features with the biased CVs. The correlation will vary greatly, also within the selected feature set. In Figure~\ref{fig:ala-wt-sampling}, we show the marginal distributions for two examples from the feature set. For a feature correlated with the biased CVs, the biasing enhances the fluctuations, and we observe a significant difference between the equilibrium distribution and the biased one, as expected. In this case, the effect of introducing $\alpha$ is to interpolate between these two limits. On the other hand, for a feature not correlated to the biased CVs, the equilibrium and biased distribution are almost the same, and $\alpha$ does not affect the distribution of this feature.

\begin{figure}[htp]
  \includegraphics[width=0.9\columnwidth]{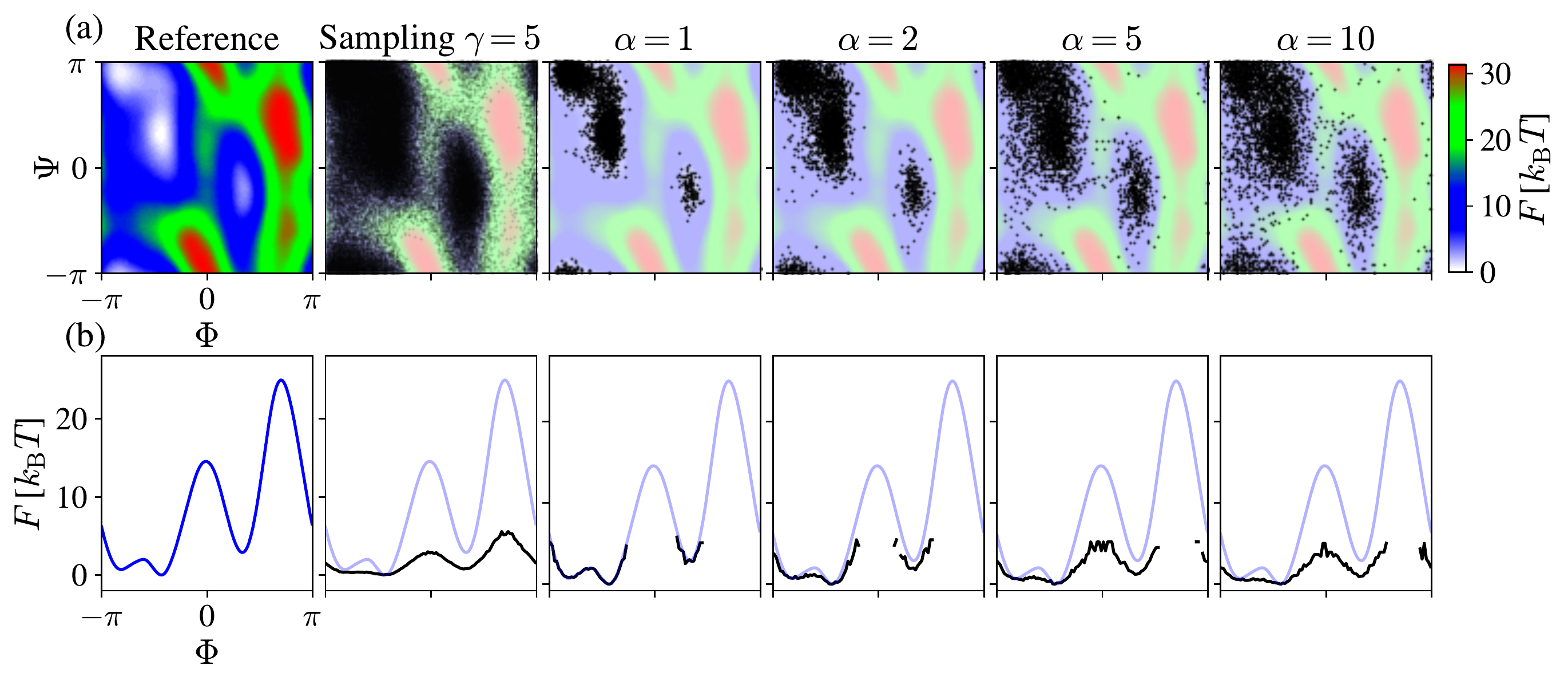}
  \caption{Results for alanine dipeptide in vacuum at 300 K. Weight-tempered random sampling as a landmark selection scheme for a WT-MetaD simulation ($\gamma=5$) biasing $(\Phi, \Psi)$. (a) In the first two panels, we show the reference FES in the $(\Phi, \Psi)$ space and the points sampled during the simulations. In the subsequent panels, we present the 4000 landmarks selected for different values of the $\alpha$ parameter. (b) In the bottom row, we show the results projected on $\Phi$, where the reference FES is shown in light blue. The projections (black) are calculated as a negative logarithm of the histogram of the selected landmarks.}
  \label{fig:ala-weights}
\end{figure}

In Figure~\ref{fig:ala-weights}, we show the results from the landmark selection for one of the WT-MetaD simulations ($\gamma=5$). In the top row, we show how the selected landmarks are distributed in the CV space. In the bottom row, we show the effective FES of selected landmarks projected on the $\Psi$ dihedral angle.

For $\alpha=1$, equivalent to weighted random sampling~\cite{tribello2019using_b}, we can see that we get a worse representation of the C$7_{\mathrm{ax}}$ state as compared to the other states. We can understand this issue by considering the weights of configurations in the C$7_{\mathrm{ax}}$ that are are considerably smaller than the weights from the other states. As shown in Section~S10 in the SI, using the $\alpha=1$ landmarks results in an MRSE embedding close to the equilibrium PT embedding (shown in Figure~\ref{fig:ala-embeddings}(a) below), but has a worse separation of the metastable states as compared to other embeddings.

On the other hand, if we use $\alpha=2$, we obtain a much more balanced landmark selection that is relatively close to the equilibrium distribution but has a sufficient representation of the C$7_{\mathrm{ax}}$ state. Using larger values of $\alpha$ renders a selection closer to the sampling from the underlying biased simulation, with more features higher in free energy. We observe that using $\alpha=2$ gives the best MRSE embedding. In contrast, higher values of $\alpha$ result in worse embeddings characterized by an inadequate mapping of the C$7_{\mathrm{ax}}$ state, as can be seen in Section~S12 in the SI. Therefore, in the following, we use a value of $\alpha=2$ for the tempering parameter in the landmark selection. This value corresponds to an effective landmark CV distribution broadening of $\tilde{\alpha} \approx 1.67$ (see eqs~\ref{eq:effective_alpha_pdf} and~\ref{eq:effective_alpha}).

These landmark selection results underline the importance of having a balanced selection of landmarks that is close to the equilibrium distribution and gives a proper representation of all metastable states, but excludes points from unimportant higher-lying free energy regions.  The exact value of $\alpha$ that achieves such optimal selection will depend on the underlying free energy landscape.

In Section~S11 in the SI, we show results obtained using WT-FPS for the landmark selection (see Section~S3 in the SI for a description of WT-FPS). We can observe that the WT-MetaD embeddings obtained using WT-FPS with $\alpha=2$  are similar to the WT-MetaD embeddings shown in Figure~\ref{fig:ala-embeddings} below. Thus, for small values of the tempering parameter, both methods give similar results.

Having established how to perform the landmark selection, we now consider the results for MRSE embeddings obtained on unbiased and biased simulation data at 300 K. The unbiased simulation data comes from a PT simulation that accurately captures the equilibrium distribution within each replica~\cite{parallel_tempering}. Therefore, for the 300 K replica used for the analysis and training, we obtain the equilibrium populations of the different metastable states while not capturing the higher-lying and transition state regions. In principle, we could also include simulation data from the higher-lying replica into the training by considering statistical weights to account for the temperature difference, but this would defeat the purpose of using the PT to generate unbiased simulation data that does not require reweighting. We refer to the embedding trained on the PT simulation data as the PT embedding. The biased simulation data comes from WT-MetaD simulations where we bias the ($\Phi$, $\Psi$) angles. We refer to these embeddings as the WT-MetaD embeddings.

In the WT-MetaD simulations, we use bias factors from 2 to 10 to generate training data sets representing a biased distribution that progressively goes from a distribution closer to the equilibrium one to more flatter distribution as we increase $\gamma$ (see eq~\ref{eq:wt-pv}). In this way, we can test how the MRSE training and reweighting procedure works when handling simulation data obtained under different biasing strengths.

For the WT-MetaD training data sets, we also investigate the effect of not incorporating the weight into the training via a reweighted feature pairwise probability distribution (i.e., all weights equal to unity in eq~\ref{eq:reweighted_kernel}). In this case, only the weight-tempered random sampling landmark selection takes the weights into account. In the following, we refer to these WT-MetaD embeddings as without reweighting or not-reweighted.

To be consistent and allow for a fair comparison between embeddings, we evaluate all the trained WT-MetaD embeddings on the unbiased PT simulation data and use the resulting projections to perform analysis and generate FESs. This procedure is possible as both the unbiased PT and the biased WT-MetaD simulations sample all metastable states of alanine dipeptide (i.e., the WT-MetaD simulations do not sample metastable states that the PT simulation does not).

\begin{figure}[htp]
  \includegraphics[width=0.9\columnwidth]{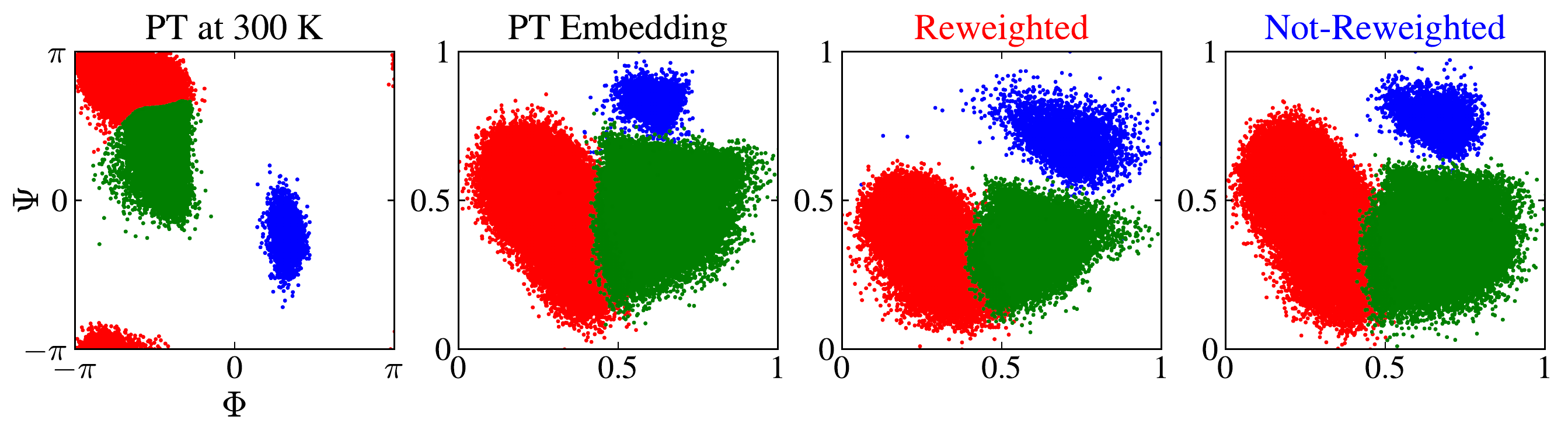}
  \caption{Results for alanine dipeptide in vacuum at 300 K. Clustering of the PT simulation data for the different embeddings. The results show how the embeddings map the metastable states. The data points are colored accordingly to their cluster. The first panel shows the metastable state clusters in the $(\Phi,\Psi)$ space. The second panel shows the results for the PT embedding. The third and fourth panels show the results for a representative case of a WT-MetaD embedding ($\gamma=5$), obtained with and without incorporating weights into the training via a reweighted feature probability distribution (see eq~\ref{eq:reweighted_kernel}), respectively. For the details about clustering~\cite{scikit-learn}, see Section~S5 in the SI. The units for the MRSE embeddings are arbitrary and only shown as a visual guide.}
  \label{fig:ala-cluster}
\end{figure}

To establish that the MRSE embeddings correctly map the metastable states, we start by considering the clustering results in Figure~\ref{fig:ala-cluster}. We can see that the PT embedding (second panel) preserves the topography of the FES and correctly maps all the important metastable states. We can say the same for the reweighted (third panel) and not-reweighted (fourth panel) embeddings. Thus, the embeddings map both the local and global characteristics of the FES accurately. Next, we consider the MRSE embeddings for the different bias factors.

\begin{figure}[htp]
  \includegraphics[width=0.8\linewidth]{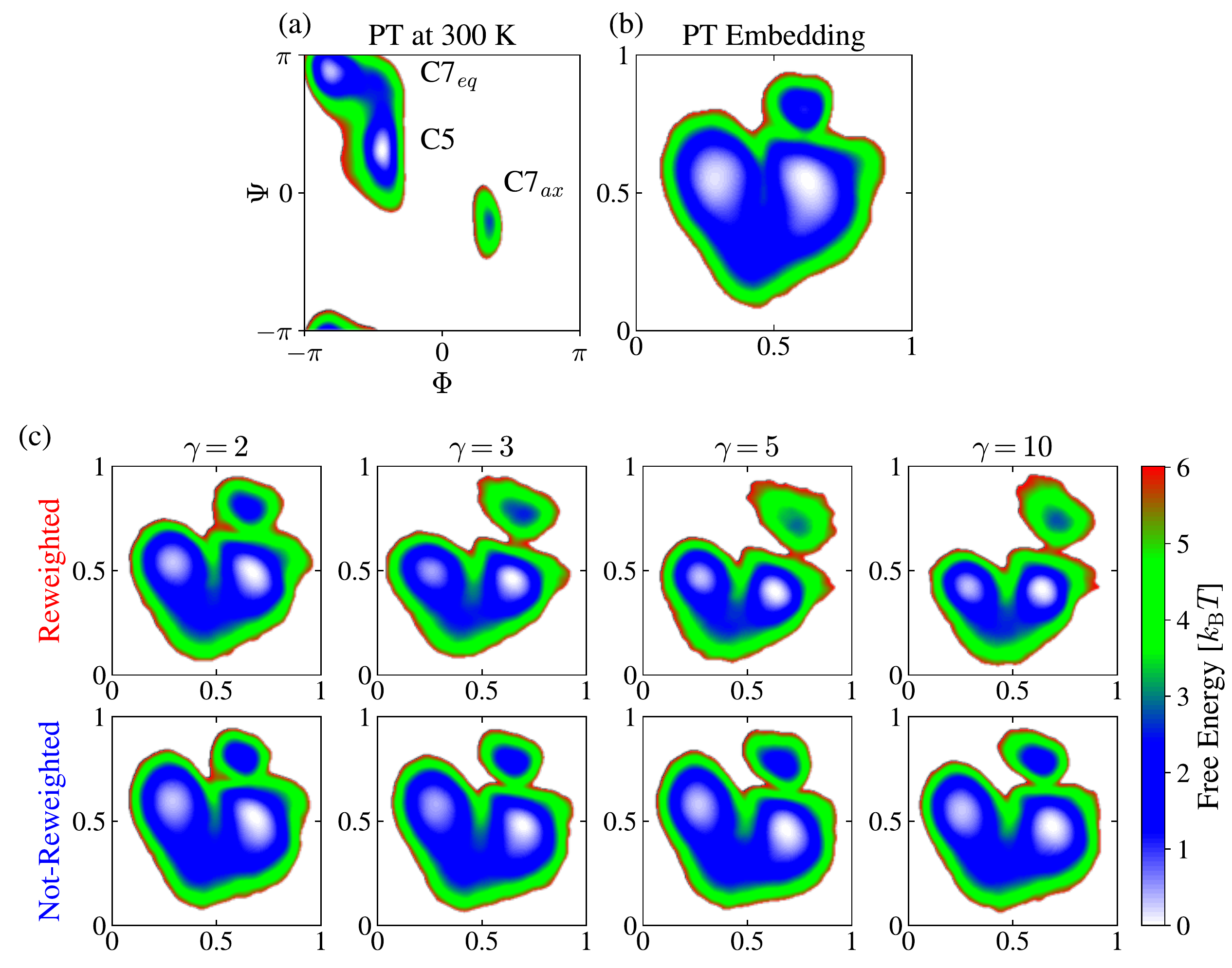}
  \caption{Results for alanine dipeptide in vacuum at 300 K. MRSE embeddings trained on unbiased and biased simulation data. (a) The free energy landscape $F(\Phi,\Psi)$ from the PT simulation. The metastable states C$7_{\mathrm{eq}}$, C5, and C$7_{\mathrm{ax}}$ are shown. (b) The FES for the MRSE embedding trained using the PT simulation data. (c) The FESs for the MRSE embeddings trained using the WT-MetaD simulation data. We show results obtained from the simulations using different bias factors $\gamma$. We show WT-MetaD embeddings obtained with (top row) and without (bottom row) incorporating weights into the training via a reweighted feature pairwise probability distribution (see eq~\ref{eq:reweighted_kernel}). We obtain all the FESs by calculating the embeddings on the PT simulation data and using kernel density estimation as described in Section~\ref{sec:kde}. The units for the MRSE embeddings are arbitrary and only shown as a visual guide.}
  \label{fig:ala-embeddings}
\end{figure}

In Figure~\ref{fig:ala-embeddings}, we show the FESs for the different embeddings along with the FES for the $\Phi$ and $\Psi$ dihedral angles. For the reweighted WT-MetaD embeddings (top row of panel c), we can observe that all the embeddings are of consistent quality and exhibit a clear separation of the metastable states. In contrast, we can see that the not-reweighted WT-MetaD embeddings (bottom row of panel c) have a slightly worse separation of the metastable states. Thus, we can conclude that incorporating the weights into the training via a reweighted feature pairwise probability distribution (see eq~\ref{eq:reweighted_kernel}) improves the visual quality of the embeddings for this system.

\begin{figure}[htp]
  \includegraphics[width=0.5\columnwidth]{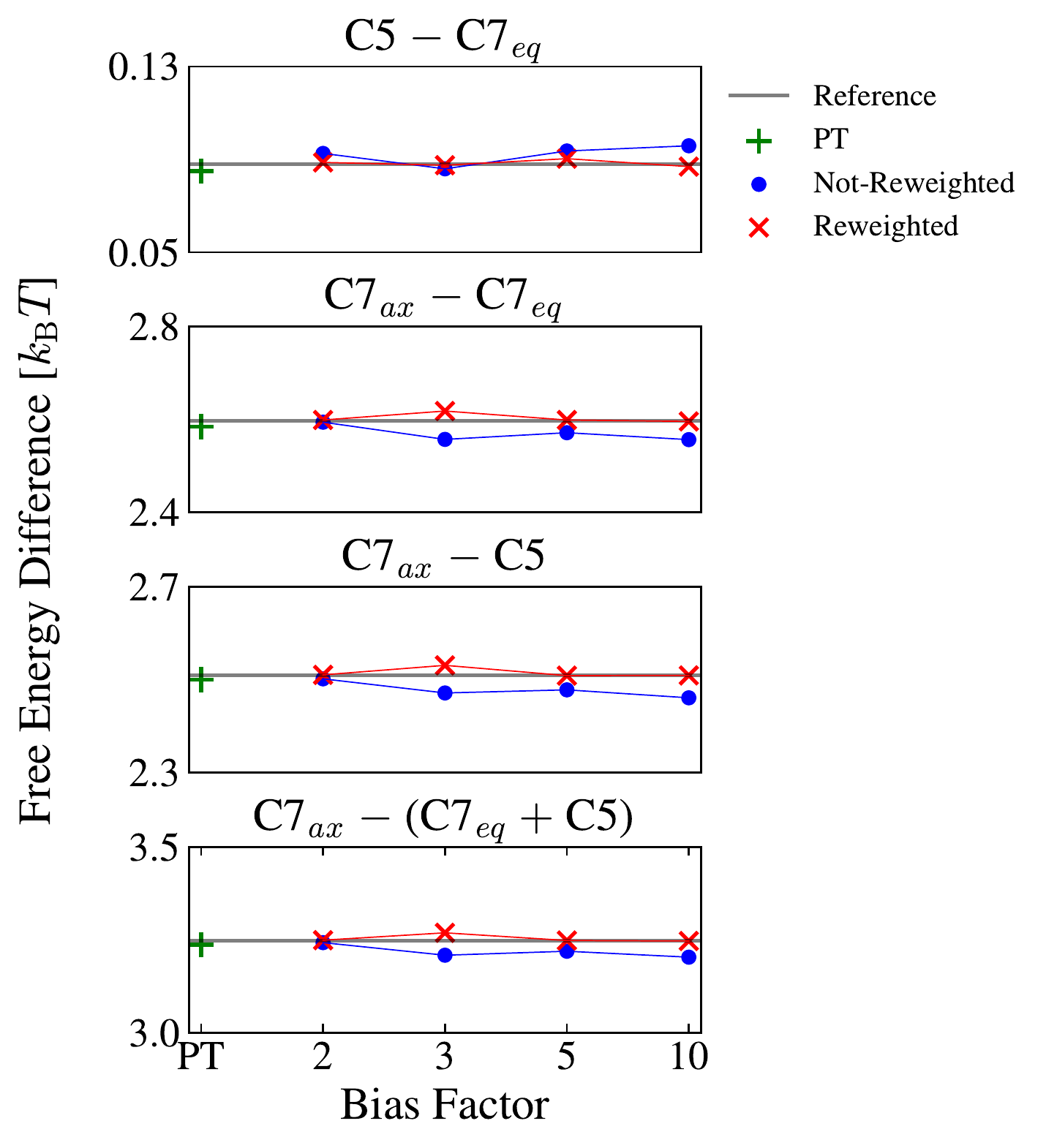}
  \caption{Results for alanine dipeptide in vacuum at 300 K. Free energy differences between metastable states for the FESs of the embeddings shown in Figure~\ref{fig:ala-embeddings}. We show the reference values from the $F(\Phi, \Psi)$ FES obtained from the PT simulation at 300 K as horizontal gray lines. The results for the reweighted embeddings are shown as red crosses, while the results for the not-reweighted embeddings are shown as blue dots. The results for the PT embedding are shown as green plus symbols.}
  \label{fig:ala-fed}
\end{figure}

To further check the quality of the embeddings, we calculate the free energy difference between metastable states as $\Delta F_{\text{A,B}}=-\frac{1}{\beta}\log(\int_{\text{A}}\d\bs\,\e^{-\beta F(\bs)} / \int_{\text{B}}\d\bs\,\e^{-\beta F(\bs)})$, where the integration domains are the regions in CV space corresponding to the states A and B, respectively. This equation is only valid if the CVs correctly discriminate between the different metastable states. For the MRSE embeddings, we can thus identify the integration regions for the different metastable states in the FES and calculate the free energy differences. Reference values can be obtained by integrating the $F(\Phi,\Psi)$ FES from the PT simulation. A deviation from a reference value would indicate that an embedding does not correctly map the density of the metastable states. In Figure~\ref{fig:ala-fed}, we show the free energy differences for all the MRSE embeddings. All free energy differences obtained with the MRSE embeddings agree with the reference values within a 0.1 $\kT$ difference for both reweighted and not-reweighted WT-MetaD embeddings. For bias factors larger than 3, we can observe that the reweighted embeddings perform distinctly better than the not-reweighted ones.

\begin{figure}[htp]
  \includegraphics[width=\linewidth]{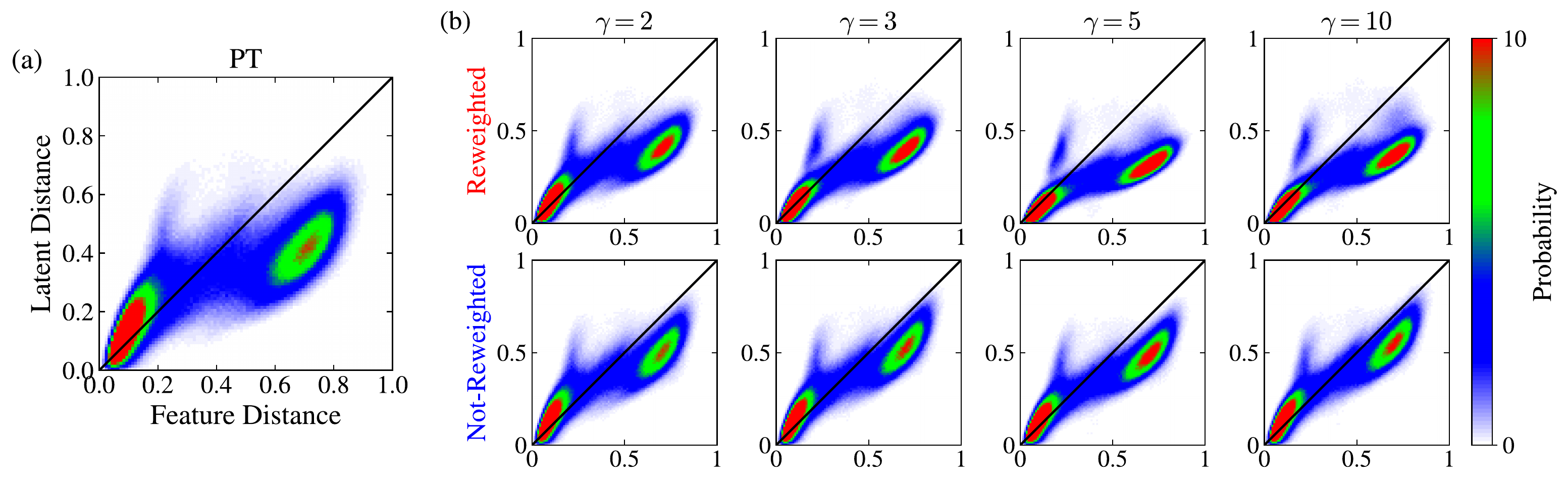}
  \caption{Results for alanine dipeptide in vacuum at 300 K. The joint probability density functions for the pairwise distances in the high-dimensional feature space and the low-dimensional latent space for the embeddings shown in Figure~\ref{fig:ala-embeddings}. We show the results for the (a) PT and (b) WT-MetaD embeddings (evaluated on the PT simulation data). These histograms show the similarities between distances in the feature and latent spaces. For an embedding that preserves distances accurately, the density would lie on the identity line $y=x$ (shown as a black line). We normalize the distances to lie in the range 0 to 1.}
  \label{fig:ala-distances}
\end{figure}

As a final test of the MRSE embeddings for this system, we follow the approach used by Tribello and Gasparotto~\cite{tribello2019using_a,tribello2019using_b}. We calculate the pairwise distances between points in the high-dimensional feature space and the corresponding pairwise distances between points in the low-dimensional latent (i.e., CV) space given by the embeddings. We then calculate the joint probability density function of the distances using histogramming. The joint probability density should be concentrated on the identity line if an embedding preserves distances accurately. However, this only is valid for the MRSE embeddings constructed without incorporating the weights into the training, since for this case, there are no additional constraints besides geometry.

As we can see in Figure~\ref{fig:ala-distances}, the joint density is concentrated close to the identity line for most cases. For the reweighted WT-MetaD embeddings (panel b), the density for the distances in the middle range slightly deviates from the identity line in contrast to the not-reweighted embeddings. This deviation is due to additional constraints on the latent space. In the reweighted cases, apart from the Euclidean distances, we also include the statistical weights into the construction of the feature pairwise probability distribution. Consequently, having landmarks with low weights in the feature space decreases the probability of being neighbors to these landmarks in the latent space. Therefore, the deviation from the identity line must be higher for the reweighted embeddings.

Summarizing the results in this section, we can observe that MRSE can construct embeddings, both from unbiased and biased simulation data, that correctly describe the local and global characteristics of the free energy landscape of alanine dipeptide. For the biased WT-MetaD simulation data, we have investigated the effect of not including the weights in the training of the MRSE embeddings. Then only the landmark selection takes the weights into account. The not-reweighted embeddings are similar or slightly worse than the reweighted ones. We can explain the slight difference between the reweighted and not-reweighted embeddings by that the weight-tempered random sampling does the primary reweighting. Nevertheless, we can conclude that incorporating the weights into the training is beneficial for the alanine dipeptide test case.

\subsection{Alanine Tetrapeptide}
\label{sec:ala3_results}
As the last example, we consider alanine tetrapeptide, a commonly used test system for enhanced sampling methods~\cite{valsson2015well,tiwary2016spectral,mccarty2017variational,yang2018refining,bonati2019neural,Invernizzi2020opus,gilberti2020atlas}. Alanine tetrapeptide is a considerably more challenging test case than alanine dipeptide. Its free energy landscape consists of many metastable states, most of which are high in free energy and thus difficult to capture in an unbiased simulation. We anticipate that we can only obtain an embedding that accurately separates all of the metastable states by using training data from an enhanced sampling simulation, which better captures higher-lying metastable states. Thus, the system is a good test case to evaluate the performance of the MRSE method and the reweighting procedure.

\begin{figure}[htp]
  \includegraphics[width=\columnwidth]{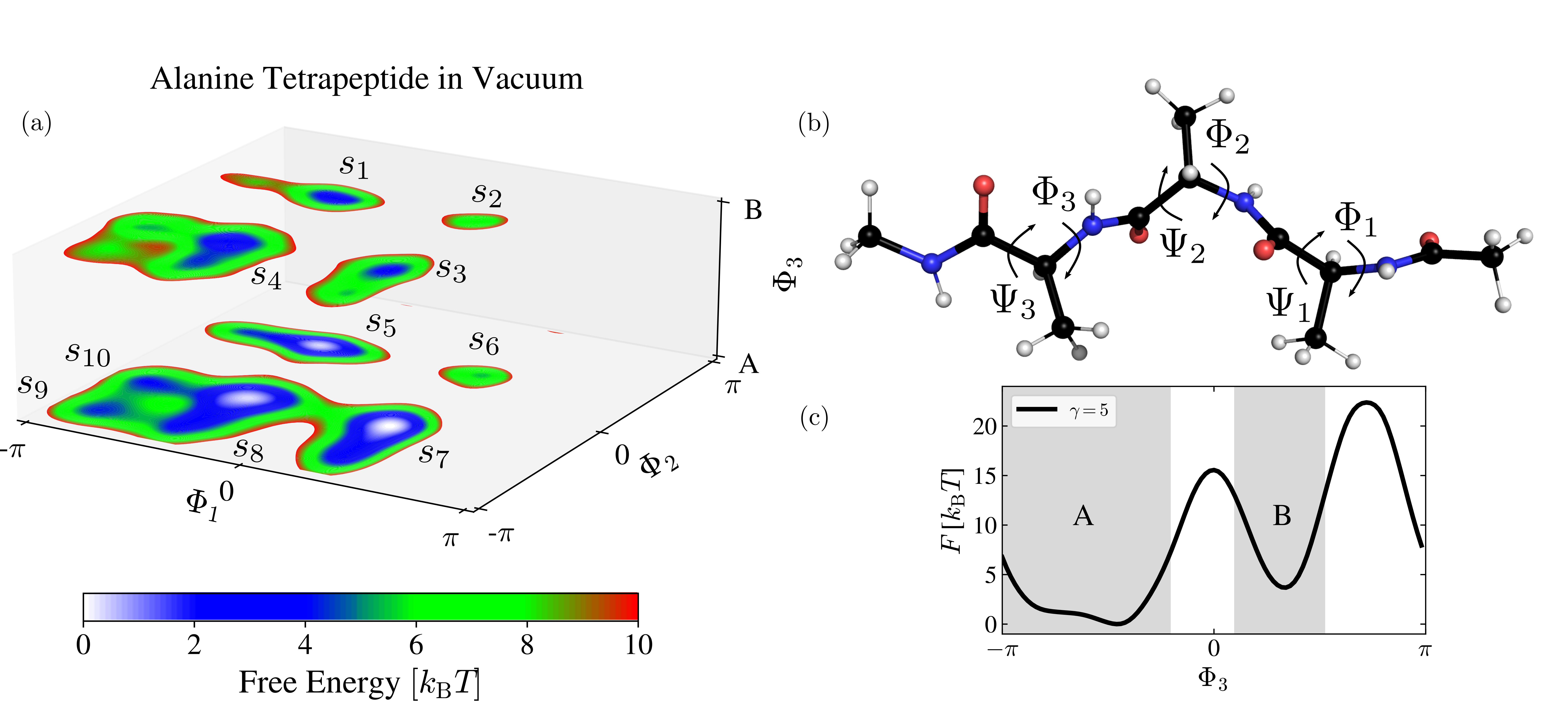}
  \caption{Results for alanine tetrapeptide in vacuum at 300 K. (a) The conditional FESs (eq~\ref{eq:conditional_fe}), obtained from the WT-MetaD simulation, shown as a function of $\Phi_1$ and $\Phi_2$ for two minima of $\Phi_3$ labeled as A and B. We denote the ten metastable states as $s_1$ to $s_{10}$. (b) The alanine tetrapeptide system with the backbone dihedral angles $\mathbf{\Phi} \equiv (\Phi_1,\Phi_2,\Phi_3)$ and $\mathbf{\Psi} \equiv (\Psi_1,\Psi_2,\Psi_3)$ that we use as the input features for the MRSE embeddings. (c) The free energy profile $F(\Phi_3)$, obtained from the WT-MetaD simulation, with the two minima A and B. The grey shaded area indicates the areas integrated over in eq~\ref{eq:conditional_fe}. The FESs are obtained using kernel density estimation as described in Section~\ref{sec:kde}.}
  \label{fig:ala3-system}
\end{figure}

As it is often customary~\cite{valsson2015well,tiwary2016spectral,Invernizzi2020opus,gilberti2020atlas}, we consider the backbone dihedral angles $\mathbf{\Phi} \equiv (\Phi_1,\Phi_2,\Phi_3)$ and $\mathbf{\Psi} \equiv (\Psi_1,\Psi_2,\Psi_3)$ that characterize the configurational landscape of alanine tetrapeptide. We show the dihedral angles in Figure~\ref{fig:ala3-system}(b). For this particular setup in vacuum, it is sufficient to use $\mathbf{\Phi}$ to describe the free energy landscape and separate the metastable states, as $\mathbf{\Psi}$ are fast CVs in comparison to $\mathbf{\Phi}$~\cite{valsson2015well,Invernizzi2020opus}. To generate biased simulation data, we perform a WT-MetaD simulation using the $\mathbf{\Phi}$ angles as CVs and a bias factor $\gamma=5$. Moreover, we perform a PT simulation and employ the 300 K replica to obtain unbiased simulation data. As before, the embeddings obtained by training on these simulation data sets are denoted as WT-MetaD and PT embeddings, respectively. As before, we also consider a WT-MetaD embedding, denoted as not-reweighted, where we do not include the weights into the construction of the feature pairwise probability distribution.

To verify the quality of the sampling and the accuracy of the FESs, we compare the results obtained from the WT-MetaD and PT simulations to results from bias-exchange metadynamics simulations~\cite{piana2007bias} using $\mathbf{\Phi}$ and $\mathbf{\Psi}$ as CVs (see Section~S13 in the SI). Comparing the free energy profiles for $\mathbf{\Phi}$ obtained with different methods (Figure~S12 in the SI), and keeping in mind that the 300 K replica from the PT simulation only describes well the lower-lying metastable states, we find that all simulations are in good agreement. Therefore, we conclude that the WT-MetaD and PT simulations are converged.

To show the results from the three-dimensional CV space on a two-dimensional surface, we consider a conditional FES where the landscape is given as a function of $\Phi_1$ and $\Phi_2$ conditioned on values of $\Phi_3$ being in one of the two distinct minima shown in Figure~\ref{fig:ala3-system}(c). We label these minima as A and B. We define the conditional FES as:
\begin{equation}
  \label{eq:conditional_fe}
  F(\Phi_1,\Phi_2 | \Phi_3 \in S)=-\frac{1}{\beta}\log\int_{S} \mathrm{d}\Phi_3 \, \e^{-\beta F(\mathbf{\Phi})},
\end{equation}
where $F(\mathbf{\Phi})$ is the FES obtained from the WT-MetaD simulation (aligned such that its minimum is at zero), $S$ is either the A or B minima, and we integrate over the regions indicated by the gray areas in Figure~\ref{fig:ala3-system}(c). We show the two conditional FESs in Figure~\ref{fig:ala3-system}(a). Through a visual inspection of Figure~\ref{fig:ala3-system}, we can identify ten different metastable states, denoted as $s_1$ to $s_{10}$. Three of the states, $s_{5}$, $s_{7}$, and $s_{8}$, are sampled properly in the 300 K replica of the PT simulation, and thus we consider them as the equilibrium metastable states. The rest of the metastable states are located higher in free energy and only sampled accurately in the WT-MetaD simulation. The number of the metastable states observed in Figure~\ref{fig:ala3-system}(a) is in agreement with a recent study of Giberti et al.~\cite{gilberti2020atlas}.

We can judge the quality of the MRSE embeddings based on whether they can correctly capture the metastable states in only two dimensions. As input features for the MRSE embeddings, we use sines and cosines of backbone dihedral angles $\mathbf{\Phi}$ and $\mathbf{\Psi}$ (12 features in total), instead of heavy atom distances as we do in the previous section for alanine dipeptide. We use weight-tempered random sampling with $\alpha=2$ to select landmarks for the training of the WT-MetaD embeddings.

\begin{figure}[htp]
  \includegraphics[width=0.9\columnwidth]{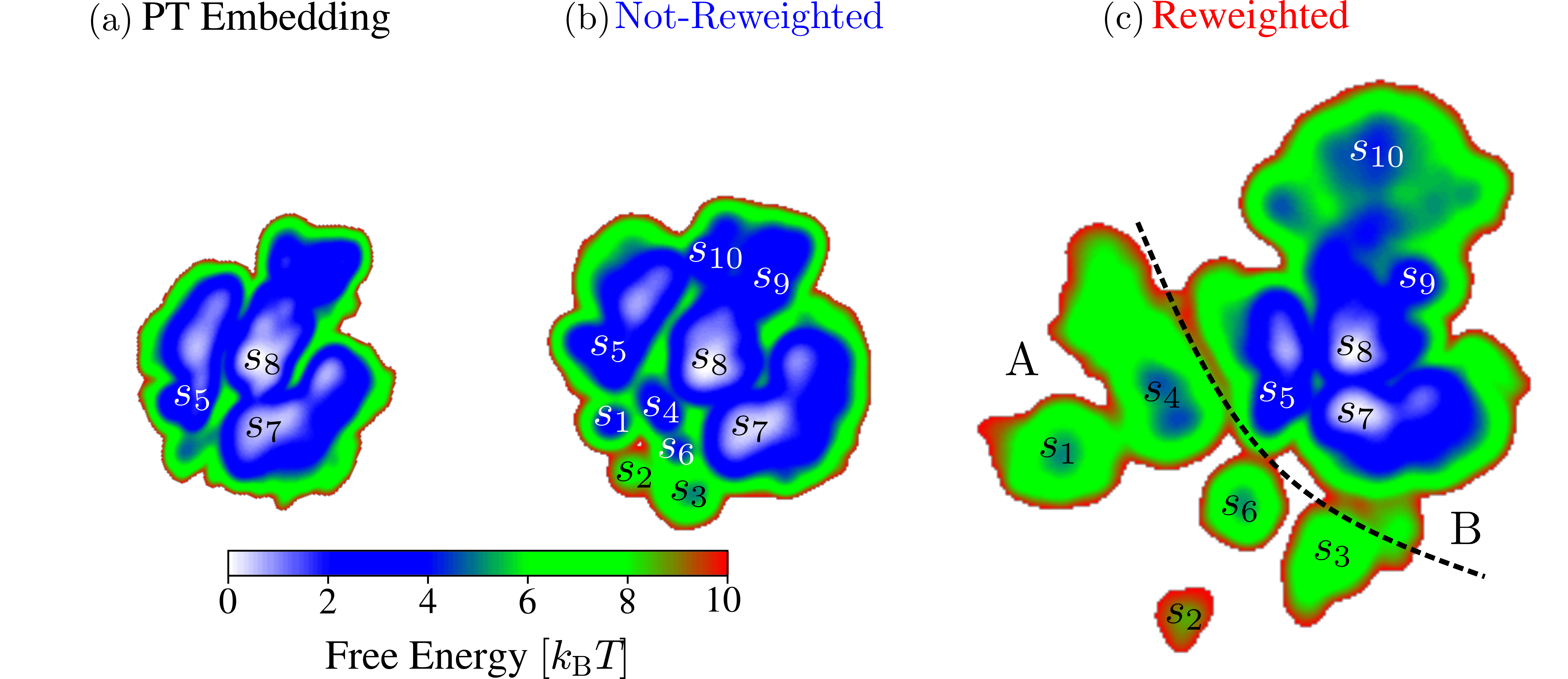}
  \caption{Results for alanine tetrapeptide in vacuum at 300 K. FESs for the MRSE embeddings trained on the unbiased and biased simulation data. (a) The PT embedding trained and evaluated on the PT simulation data. (b-c) The WT-MetaD embeddings trained and evaluated on the WT-MetaD simulation data. The WT-MetaD embeddings are obtained without (b) and with (c) incorporating weights into the training via a reweighted feature pairwise probability distribution (see eq~\ref{eq:reweighted_kernel}). The FESs are obtained using kernel density estimation as described in Section~\ref{sec:kde}. The state labels in the FESs correspond to the labeling used in Figure~\ref{fig:ala3-system}(a). The embeddings are rescaled so that the equilibrium states are of similar size. The units for the MRSE embeddings are arbitrary and thus not shown.}
  \label{fig:ala3-emb}
\end{figure}

We show the PT and WT-MetaD embeddings in Figure~\ref{fig:ala3-emb}. We can see that the PT embedding in Figure~\ref{fig:ala3-emb}(a) is able to accurately describe the equilibrium metastable states (i.e., $s_5$, $s_7$, and $s_8$). However, as expected, the PT embedding cannot describe all ten metastable states, as the 300 K replica in the PT simulation rarely samples the higher-lying states.

In contrast, we can see that the WT-MetaD embeddings in Figure~\ref{fig:ala3-emb}(b-c) capture accurately all ten metastable states. By visual inspection of the simulation data, we can assign state labels for the embeddings in Figure~\ref{fig:ala3-emb}, corresponding to the states labeled in Figure~\ref{fig:ala3-system}(a). One interesting aspect of the MRSE embeddings in Figure~\ref{fig:ala3-emb} is that they similarly map the equilibrium states, even if we obtain the embeddings from different simulation data sets (PT and WT-MetaD). This similarity underlines the consistency of our approach. The fact that both the reweighted and not-reweighted WT-MetaD embeddings capture all ten states suggests we could use both embeddings as CVs for biasing.

However, we can observe that the reweighted embedding has a better visual separation of the states. For example, we can see this for the separation between $s_9$ and $s_{10}$. Furthermore, we can see that the reweighted embedding separates the states from the A and B regions better than the not-reweighted embedding. In the reweighted embedding, states $s_1$ to $s_4$ are close to each other and separated from states $s_5$--$s_{10}$ as indicated by line drawn in Figure~\ref{fig:ala3-emb}(c). Therefore, we can conclude that the reweighted WT-MetaD embedding is of better quality and better represents distances between metastable states for this system. These results show that we need to employ a reweighted feature pairwise probability distribution for more complex systems.

\section{Discussion and Conclusions}
\label{sec:discussion_conclusions}
We present multiscale reweighted stochastic embedding, a general framework that unifies enhanced sampling and machine learning for constructing collective variables. MRSE builds on top of ideas from stochastic neighbor embedding methods~\cite{hinton2002stochastic,maaten2008visualizing,maaten2009learning,van2014accelerating}. We introduce several advancements to SNE methods that make MRSE suitable for constructing CVs from biased data obtained from enhanced sampling simulations.

We show that this method can construct CVs automatically by learning a mapping from a high-dimensional feature space to a low-dimensional latent space via a deep neural network. We can use the trained NN to project any given point in feature space to CV space without rerunning the training procedure. Furthermore, we can obtain the derivatives of the learned CVs with respect to the input features and bias the CVs within an enhanced sampling simulation. In future work, we will use this property by employing MRSE within an enhanced sampling scheme where the CVs are iteratively improved~\cite{zhang2018unfolding,chen2018collective,ribeiro2018reweighted}.

In this work, we focus entirely on the training of the embeddings, using training data sets obtained from both unbiased simulation and biased simulation employing different biasing strengths (i.e., bias factors in WT-MetaD). As the ``garbage in, garbage out'' adage applies to ML (a model is only as good as training data), to eliminate the influence of incomplete sampling, we employ idealistic sampling conditions that are not always achievable in practice~\cite{pant2020statistical}. In future work, we will need to consider how MRSE performs under less ideal sampling conditions. One possible option to address this issue is to generate multiple embeddings by running independent training attempts and score them using the maximum caliber principle, as suggested in ref~\citenum{pant2020statistical}.

The choice of the input features depends on the physical system under study. In this work, we use conventional features, i.e., microscopic coordinates, distances, and dihedral angles, as they are a natural choice for the model systems considered here. In general, the features can be complicated functions of the microscopic coordinates~\cite{musil2021physicsinspired}. For example, symmetry functions have been used as input features in studies of phase transformations in crystalline systems~\cite{geiger2013neural,rogal2019neural}. Additionally, features may be correlated or simply redundant. See ref~\citenum{dy2004feature} for a general outline of feature selection in unsupervised learning. We will explore the usage of more intricate input features and modern feature selection methods~\cite{ravindra2020automatic,cersonsky2020improving} for MRSE embeddings in future work.

One of the issues with using kernel-based dimensionality reduction methods, such as diffusion maps~\cite{coifman2008diffusion} or SNE methods~\cite{hinton2002stochastic}, is that the user needs to select the bandwidths (i.e., the scale parameters $\boldsymbol{\varepsilon}$) when using the Gaussian kernels. In $t$-SNE~\cite{maaten2008visualizing,maaten2009learning}, the Gaussian bandwidths are optimized by fitting to a parameter called perplexity. We can view the perplexity as the effective number of neighbors in a manifold~\cite{maaten2008visualizing,maaten2009learning}. However, this only redirects the issue as the user still needs to select the perplexity parameter~\cite{wattenberg2016how}. Larger perplexity values lead to a larger number of nearest neighbors and an embedding less sensitive to small topographic structures in the data. Conversely, lower perplexity values lead to fewer neighbors and ignore global information in favor of the local environment. However, what if several length scales characterize the data? In this case, it is impossible to represent the density of the data with a single set of bandwidths, so viewing multiple embeddings obtained with different perplexity values is quite common~\cite{wattenberg2016how}.

In MRSE, we circumvent the issue of selecting the Gaussian bandwidths or the perplexity value by employing a multiscale representation of feature space. Instead of a single Gaussian kernel, we use a Gaussian mixture where each term has its bandwidths optimized for a different perplexity value. We perform this procedure in an automated way by employing a range of perplexity values representing several length scales. This mixture representation allows describing both the local and global characteristics of the underlying data topography. The multiscale nature of MRSE makes the method particularly suitable for tackling complex systems, where the free energy landscape consists of several metastable states of different sizes and shapes. However, as we have seen in Section~\ref{sec:ala3_results}, also model systems may exhibit such complex behavior.

Employing nonlinear dimensionality reduction methods is particularly problematic when considering training data obtained from enhanced sampling simulations. In this case, the feature samples are drawn from a biased probability distribution, and each feature sample carries a statistical weight that we need to take into account. In MRSE, we take the weights into account when selecting the representative feature samples (i.e., landmarks) for the training. For this, we introduce a weight-tempered selection scheme that allows us to obtain landmarks that strike a balance between equilibrium distribution and capturing important metastable states lying higher in free energy. This weight-tempered random sampling method depends on a tempering parameter $\alpha$ that allows us to tune between obtaining equilibrium and biased distribution of landmarks. This parameter is case-dependent and similar in spirit to the bias factor $\gamma$ in WT-MetaD. Generally, $\alpha$ should be selected so that every crucial metastable state is densely populated. However, $\alpha$ should not be too large, as it may result in including feature samples from unimportant higher-lying free energy regions.

The weight-tempered random sampling algorithm is inspired by and bears a close resemblance to the well-tempered farthest-point sampling (WT-FPS) landmark selection algorithm, introduced by Ceriotti et al.~\cite{ceriotti2013demonstrating}. For small values of the tempering parameter $\alpha$, both methods give similar results as discussed in Section~\ref{sec:ala1_results}. The main difference between the algorithms lies in the limit $\alpha \to \infty$. In weight-tempered random sampling, we obtain a landmark distribution that is the same as the biased distribution from the enhanced sampling simulation. On the other hand, WT-FPS results in landmarks that are sampled uniformly distributed from the simulation data set. Due to usage of FPS~\cite{Hochbaum1985_Abestpo} in the initial stage, WT-FPS is computationally more expensive. Thus, as we are interested in a landmark selection obtained using smaller values of $\alpha$ and do not want uniformly distributed landmarks, we prefer weight-tempered random sampling.

The landmarks obtained with weight-tempered random sampling still carry statistical weights that can vary considerably. Thus, we also incorporate the weights into the training by employing a reweighted feature pairwise probability distribution. To test the effect of this reweighting, we constructed MRSE embeddings without including the weights in the training. Then, we only take the weights into account during the landmark selection. For alanine dipeptide, the reweighted MRSE embeddings are more consistent and slightly better than the not-reweighted ones. For the more challenging alanine tetrapeptide case, both the reweighted and not-reweighted embeddings capture all the metastable states. However, we can observe that the reweighted embedding has a better visual separation of states. Thus, we can conclude from these two systems that employing a reweighted feature pairwise probability distribution is beneficial or even essential, especially when considering more complex systems. Nevertheless, this is an issue that we need to consider further in future work.

Finally, we have implemented the MRSE method and weight-tempered random sampling in the open-source \textsc{plumed} library for enhanced sampling and free energy computation~\cite{tribello2014plumed,plumed-nest}. Having MRSE integrated into \textsc{plumed} is of significant advantage. We can use MRSE with the most popular MD codes and learn CVs in postprocessing and on the fly during a molecular simulation. Furthermore, we can employ the learned CVs with the various CV-based enhanced sampling methods implemented in \textsc{plumed}. We will make our code publicly available under an open-source license by contributing it as a module called \texttt{LowLearner} to the official \textsc{plumed} repository in the future. In the meantime, we release an initial implementation of \texttt{LowLearner} with our data. The archive of our data is openly available at Zenodo~\cite{mrse-dataset} (DOI: \href{https://zenodo.org/record/4756093}{\texttt{10.5281/zenodo.4756093}}). \textsc{plumed} input files and scripts required to replicate the results are available from the \textsc{plumed} NEST~\cite{plumed-nest} under \texttt{plumID:21.023} at \url{https://www.plumed-nest.org/eggs/21/023/}.

\section*{Acknowledgments}
We want to thank Ming Chen (UC Berkeley) and Gareth Tribello (Queen's University Belfast) for valuable discussions, and Robinson Cortes-Huerto, Oleksandra Kukharenko, and Joseph F. Rudzinski (Max Planck Institute for Polymer Research) for carefully reading over an initial draft of the manuscript. JR gratefully acknowledges financial support from the Foundation for Polish Science (FNP). We acknowledge using the MPCDF (Max Planck Computing \& Data Facility) DataShare.

\section*{Associated Content}
The Supporting Information is available free of charge at \url{https://pubs.acs.org/doi/xxx/yyy}.

\begin{adjustwidth}{2cm}{}
(S1) Entropy of the reweighted feature pairwise probability distribution;
(S2) Kullback-Leibler divergence loss for a full set of training data;
(S3) Description of well-tempered farthest-point sampling (WT-FPS);
(S4) Effective landmark CV distribution for weight-tempered random sampling;
(S5) Details about the clustering used in Figure~7.
(S6) Bandwidth values for kernel density estimation;
(S7) Loss function learning curves;
(S8) Additional embeddings for the M\"uller-Brown potential;
(S9) Feature preprocessing in the alanine dipeptide system;
(S10) Alanine dipeptide embeddings for different values of $\alpha$ in weight-tempered random sampling;
(S11) Alanine dipeptide embeddings for $\alpha=2$ in WT-FPS;
(S12) Alanine dipeptide embeddings for different random seed values;
(S13) Convergence of alanine tetrapeptide simulations;
\end{adjustwidth}

\bibliography{MRSE-2007.06377}

\end{document}